\documentclass[letterpaper,english]{emulateapj}
\usepackage{ae,aecompl}
\usepackage[T1]{fontenc}
\usepackage[latin9]{inputenc}
\usepackage{graphicx}

\makeatletter

\pdfpageheight\paperheight
\pdfpagewidth\paperwidth

\providecommand{\tabularnewline}{\\}

\slugcomment{}

\shorttitle{}

\shortauthors{}

\usepackage{enumitem}  

\setlist{nolistsep}

\usepackage{graphicx}

\usepackage{placeins}

\makeatother

\usepackage{babel}
\begin{document}

\title{Strict Upper Limits on the Carbon-to-Oxygen Ratios of Eight Hot Jupiters\\
from Self-Consistent Atmospheric Retrieval}

\author{Björn Benneke}

\affil{Division of Geological and Planetary Sciences, California Institute
of Technology, Pasadena, CA 91125, USA}

\email{bbenneke@caltech.edu}
\begin{abstract}
The elemental compositions of hot Jupiters are informative relics
of planet formation that can help us answer long-standing questions
regarding the origin and formation of giant planets. Here, I present
the main conclusions from a comprehensive atmospheric retrieval survey
of eight hot Jupiters with detectable molecular absorption in their
near-infrared transmission spectra. I analyze the eight transmission
spectra using the newly-developed, self-consistent atmospheric retrieval
framework, SCARLET. Unlike previous methods, SCARLET combines the
physical and chemical consistency of complex atmospheric models with
the statistical treatment of observational uncertainties known from
atmospheric retrieval techniques. I find that all eight hot Jupiters
consistently require carbon-to-oxygen ratios (C/O) below $\sim0.9$.
The finding of $\mathrm{C/O<0.9}$ is highly robust for HD~209458b,
WASP-12b, WASP-19b, HAT-P-1b, and XO-1b. For HD~189733b, WASP-17b,
and WASP-43b, I find that the published \textit{WFC3} transmission
spectra favor $\mathrm{C/O<0.9}$ at greater than $95\%$ confidence.
I further show that the water abundances on all eight hot Jupiters
are consistent with solar composition. The relatively small depth
of the detected water absorption features is due to the presence of
clouds, not due to a low water abundance as previously suggested for
HD209458b. The presence of a thick cloud deck is inferred for HD~209458b
and WASP-12b. HD~189733b may host a similar cloud deck, rather than
the previously suggested Rayleigh hazes, if star spots affect the
observed spectrum. The approach taken in SCARLET can be regarded as
a new pathway to interpreting spectral observations of planetary atmospheres.
In this work, including our prior knowledge of H-C-N-O chemistry enables
me to constrain the C/O ratio without detecting a single carbon-bearing
molecule.
\end{abstract}

\section{Introduction\label{sec:Introduction}}

Spectroscopy observations of hot Jupiters provide an unprecedented
opportunity to constrain the metallicities and carbon-to-oxygen ratios
(C/O) of giant planets which can provide crucial hints on the formation
of giant planets including the gas and ice giants in our solar system.
If giant planets predominately form in a quick one-step process through
gravitational instability \citep{boss_giant_1997}, their atmospheres
should have elemental abundances resembling the ones of their host
star \citep{helled_heavy-element_2009}. If, on the other hand, giant
planets form through multi-step core accretion process \citep{pollack_formation_1996},
a wide range of elemental compositions with distinct deviations from
the host star's elemental composition are expected \citep[Mordasini et al., 2015]{oberg_disk_2011,helling_disk_2014,madhusudhan_towards_2014,ali-dib_carbon-rich_2014}.
Within the core accretion model, refractory and ice-forming elements
first segregate from the gas in the protoplanetary disk through condensation.
They then evolve separately from the gas for millions of years, and
finally, some of the gas and solids are recombined in the process
of planet formation. In this multi-step process, it would be a strange
coincidence for the final planet to form with elemental compositions
identical to the one of the host star. Instead, we expect the elemental
abundances of giant exoplanets to be insightful relics of this planet
formation process.

The carbon-to-oxygen ratio (C/O), in particular, has been proposed
to be a good tracer of planet formation. Protoplanetary disk models
suggest that the C/O ratio of the gas and solids in protoplanetary
disks varies with distance from the star because of the different
condensation temperatures of $\mathrm{H_{2}O}$ and CO \citep{oberg_disk_2011}.
Detailed disk evolution models suggest that the C/O of the gas beyond
the water iceline should transition to $\mathrm{C/O\rightarrow1}$
on time scales of $\sim3\,\mathrm{Myrs}$ \citep{helling_disk_2014}.
As a result, the gas envelopes of giant planets formed beyond the
iceline may be carbon enriched with the exact C to O ratio depending
on the time and the location of the runaway gas accretion. If, on
the other hand, gas giant atmospheres are heavily polluted by the
accretion of oxygen enriched solids, the C/O ratio of giant planet
envelopes may substantially deviate from the C/O ratio in the gas
disk. Van Boekel et al. (2015) model the formation and evolution of
giant exoplanets from the formation of planetesimals to today's evolved
giant planets, keeping track of the amounts of gaseous, icy, and rocky
material accreted as the planets forms and migrates within the disk.
They conclude that giant planet formation beyond the water ice line
should always lead to $\mathrm{C/O<1}$, while giant planet formation
within the iceline can lead to either carbon-rich chemistry ($\mathrm{C/O>1}$)
or oxygen-rich chemistry ($\mathrm{C/O<1}$) depending on the carbon
abundance of refractory materials. The uncertainties in planet formation
and disk evolution models are substantial, however. Observational
constraints on the elemental abundances of giant planets are greatly
needed. 

The challenge in obtaining reliable observational constraints on the
elemental abundances lies not only in obtaining sufficiently precise
data, but also in the difficulty associated with inferring elemental
abundances representative of the planet's bulk gas envelope from spectroscopic
observations of the planet's photosphere. Despite exquisite observations,
no representative values for the C/O ratios of the solar system gas
giants, no representative values for the C/O ratio in the Solar System
giants have been determined to date, because oxygen is condensed out
in the form of water in deep layers, largely inaccessible to remote
sensing. More accurate oxygen abundances and C/O ratios for the solar
system gas giants would provide an important new constraint on our
models of gas giant planet formation. One of the main science objectives
of the Juno mission --- expected to arrive in 2016 --- is to provide
the first representative C/O measurement for Jupiter's gas envelope
through microwave observations from a low Jupiter orbit.

Observations of hot Jupiters and directly imaged young giant planets
present a new opportunity to determine the elemental abundances for
a large sample of giant planets using infrared spectroscopy. The temperatures
in their atmospheres are sufficiently high that all of the dominant
carbon or oxygen bearing molecular species are accessible to IR remote
sensing. However, the spectral signatures of exoplanets are governed
not only by the planet's elemental composition, but also its radiation
environment and a wide range of weakly understood chemical and dynamical
processes in their atmosphere.

Here, I introduce a new ``Self-Consistent Atmospheric Retrieval framework
for ExoplaneTs'' (SCARLET) which aims to provide direct constraints
on elemental abundances while accounting for the uncertainties resulting
from our limited understanding of the chemical, dynamical and cloud
formation processes in the atmospheres of these planets. Unlike previous
methods, the new framework is able to determine the full range of
self-consistent scenarios for a given planet by combining the observational
data and our prior knowledge of atmospheric chemistry and physics
in a statistically robust Bayesian analysis. One key concept is that
for a given metallicity and C/O ratio there is not a single model
transmission spectrum to be compared to the data, but instead a range
of models corresponding to the uncertainties introduced by our limited
understanding of the relevant atmospheric processes. SCARLET accounts
for this model uncertainty by marginalizing over a wide range of ``atmospheric
process parameters'' whose prior distributions quantify our limited
knowledge of vertical mixing, chemistry, and cloud formation in the
atmospheres of hot exoplanets.

In this first work, I apply the SCARLET framework to eight published
hot Jupiter transmission spectra with detectable near-IR molecular
absorption features (HD~209458b, WASP-19b, HAT-P-1b, XO-1b, HD~189733b,
WASP-12b, WASP-17b, and WASP-43b). In this study, we focus on \textit{HST
WFC3} observations because they provide substantial \textit{simultaneous}
spectral coverage ($1.1-1.7\,\mathrm{\mu m}$), and have been shown
to be extremely repeatable \citep[e.g.,][]{kreidberg_clouds_2014},
leaving little doubt about the reliability of the water detections.
Where available we include \textit{HST} \textit{STIS} and \textit{Spitzer}
observations in the analysis --- note, however, that the C/O constraints
presented in this work rely only on the \textit{HST} \textit{WFC3}
observations. Our primary conclusions do not depend on comparisons
of observations taken at different times, and the repeatability of\textit{
WFC3} observations strongly indicates that the results are not compromised
by systematic effects that may have affected early C/O measurements
based on broadband Spitzer \textit{IRAC} data points \citep[e.g., ][]{madhusudhan_high_2011,crossfield_re-evaluating_2012,cowan_thermal_2012}. 

In this article, I first motivate the development of the new SCARLET
framework by reviewing the strengths and weaknesses of previous modeling
approaches for analyzing planetary atmospheres (Section 2). Section
3 then describes the details of the SCARLET framework. Section 4 describes
the results from the retrieval study of four hot Jupiters. Section
5 presents the conclusions from this work and discusses the implications
for hot Jupiter formation and migration.

\begin{figure}[t]
\noindent \begin{centering}
\includegraphics[width=1\columnwidth]{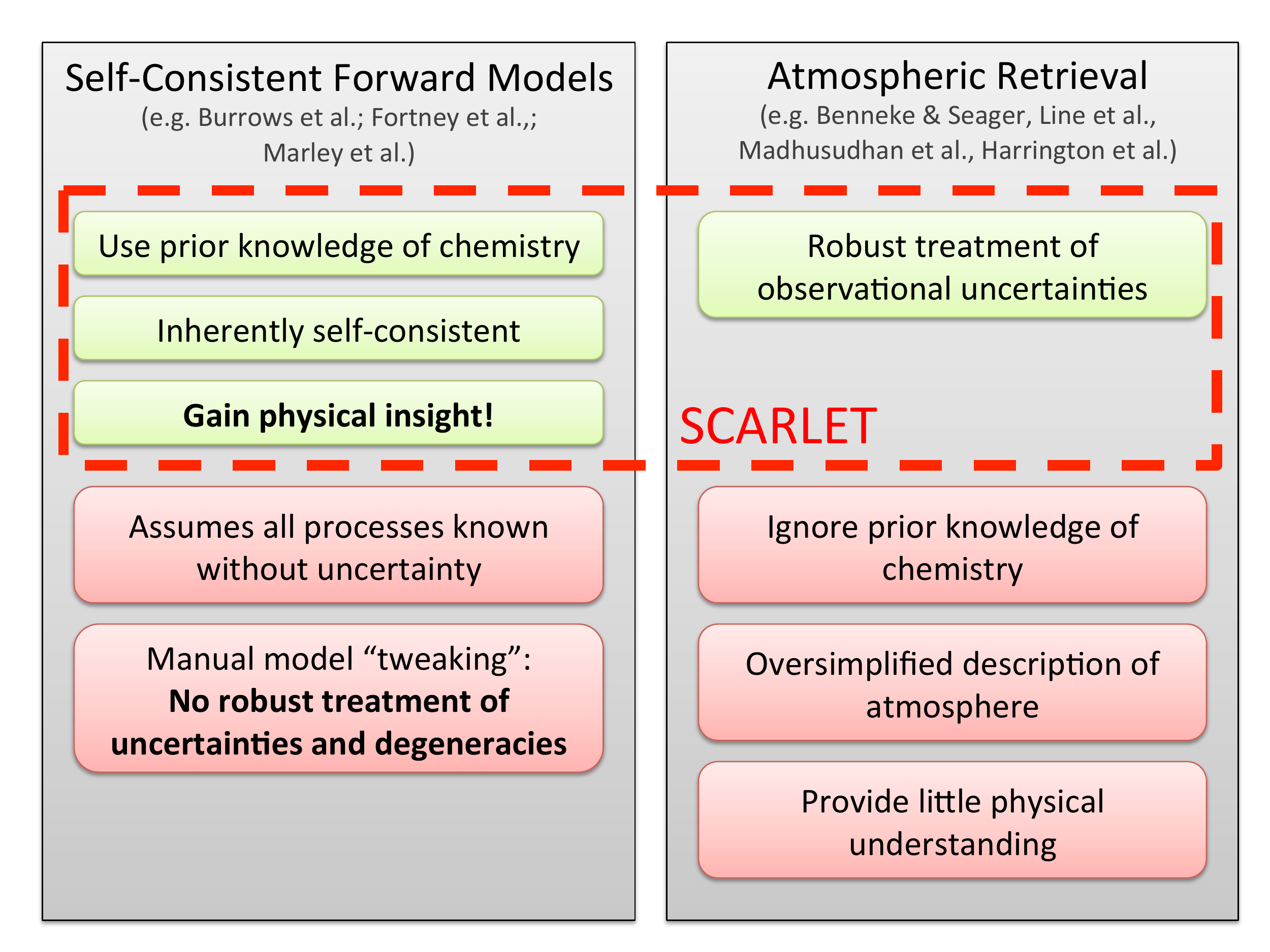}
\par\end{centering}

\protect\caption{Strength and weakness of self-consistent forward models and atmospheric
retrieval techniques for the interpretation of exoplanet spectra.
Strengths are marked in green. Weakness are marked in red. The new
``SCARLET'' approach combines the strengths of self-consistent forward
modeling and atmospheric retrieval techniques and overcomes their
weaknesses. \label{fig:Motivation-for-SCARLET}}
\end{figure}

\section{Background: Interpreting Planetary Spectra}

The interpretation of spectral observations of solar system atmospheres,
exoplanets, and brown dwarfs has historically been performed using
two distinct modeling strategies: ``Atmosphere forward modeling''
is a theory-driven approach in which the governing equations of chemistry
and physics are solved to derive a small number of self-consistent
scenarios for the atmosphere. ``Atmospheric retrieval'', on the
other hand, is an observation-driven approach in which statistically
robust constraints on the atmospheric state molecular composition
and temperature structure are retrieved by repeatedly comparing observations
to model spectra of parameterized scenarios without modeling the chemical
and physical process in the atmosphere. Both approaches have strengths
and severe limitations (Figure \ref{fig:Motivation-for-SCARLET}).
The goal of this section is to provide a concise overview of ``forward
models'' and ``atmospheric retrieval techniques'' to motivate the
development of the new SCARLET framework presented in this work.

\subsection{Complex Atmospheric ``Forward'' Models}

The basic idea of atmosphere forward models is to iterate the relevant
atmospheric processes including atmospheric chemistry, radiative-convective
heat transport, and/or cloud formation until a converged self-consistent
solution is achieved \citep{burrows_nongray_1997,seager_theoretical_2000,fortney_synthetic_2008,showman_atmospheric_2009,menou_water-trapped_2013}.
Depending on the application, particular emphasis is assigned on modeling
the detailed chemistry \citep[e.g.,][]{zahnle_atmospheric_2009,moses_disequilibrium_2011,hu_photochemistry_2012},
the atmospheric dynamics \citep[e.g.,][]{showman_atmospheric_2009},
the thermal structure \citep[e.g.,][]{mckay_thermal_1989,burrows_nongray_1997,fortney_self-consistent_2011}
and/or the formation of clouds \citep[e.g.,][]{marley_clouds_2002,lavvas_titans_2010,morley_quantitatively_2013}.
The main strength of these atmosphere ``forward'' models is that
they provide physical insights into the atmospheric processes at play.
Matching the observations generally requires modeling all relevant
physical processes. If solved iteratively, atmospheric scenarios from
atmosphere forward models are generally physically self-consistent
within the modeled physics. 

Severe limitations of atmosphere ``forward'' modeling are, however,
that they inherently assume that all physical and chemical processes
are known without uncertainty (Figure \ref{fig:Motivation-for-SCARLET}).
Mismatch between data and models is often reduced by manually tweaking
individual model parameters until a sufficient match to the data is
achieved. Manual tweaking, however, prevents a statistically robust
treatment of the observational uncertainty. Individual scenarios matching
the data are identified without understanding what the degeneracies
are and whether the found scenario is the only plausible scenario.

\subsection{Atmospheric Retrieval}

Atmospheric retrieval methods are different from atmosphere ``forward''
models in that they generally do not model any of the chemical and
dynamical processes in the atmosphere. Instead, they describe the
molecular composition and temperature profile by free parameters and
model the planet's spectrum for comparison to the data. Atmospheric
techniques are powerful when good data is available and have a long
history of applications in meteorology \citep[e.g.][]{smith_iterative_1970,chahine_remote_1974}
and the exploration of the solar system planets \citep[e.g.,][]{hanel_exploration_2003}.
An excellent overview of the theory of atmospheric retrieval within
the assumptions of Gaussian uncertainties is provided in \citet{rodgers_inverse_2000}.
Techniques that more robustly explore the full parameter space of
planetary atmospheres are described in \citet{madhusudhan_temperature_2009,benneke_atmospheric_2012,line_information_2012}. 

A key disadvantages of atmospheric retrieval techniques is that they
provide no direct insights into the physical and chemical processes
at play. They also make no use of our prior substantial knowledge
of atmospheric chemistry and heat transport to better constrain the
properties of the planet studied.  This is particularly critical
for exoplanets because the available data is sparse and including
our prior knowledge of chemistry or heat transport can substantially
reduce the uncertainties in the inferred elemental abundances and
internal heat flux. Instead, atmospheric retrieval techniques make
no attempt to rule out unphysical scenarios that grossly contradict
our basic understanding of atmospheric chemistry and radiative-convective
heat transport. Combinations of molecular gases that are highly reactive
with each other at given conditions are not ruled out in regular retrieval
methods (see \citealp{hu_photochemistry_2014} for a discussion).
Similarly,  no consistency check between the parameterized temperature-pressure
profile and the molecular composition is performed \citep{madhusudhan_temperature_2009,line_systematic_2013-1}).
Equally important, atmospheric retrieval techniques for exoplanets
require an oversimplified description of atmosphere in order to reduce
the number of free parameters. Well-mixed atmospheres composed of
only a small number of pre-selected molecules are considered, even
though the mixing ratios realistic atmospheres can be strongly altitude
dependent \citep{moses_compositional_2013,moses_chemical_2013}. The
presented shortcoming of atmospheric retrieval techniques suggest
the potential for an improved way to analyze spectroscopic observations
of exoplanets --- one that combines the statistical robustness of
atmospheric retrieval techniques with the physical insights provided
by atmosphere forward models.

\begin{figure}[t]
\noindent \centering{}\includegraphics[width=1\columnwidth]{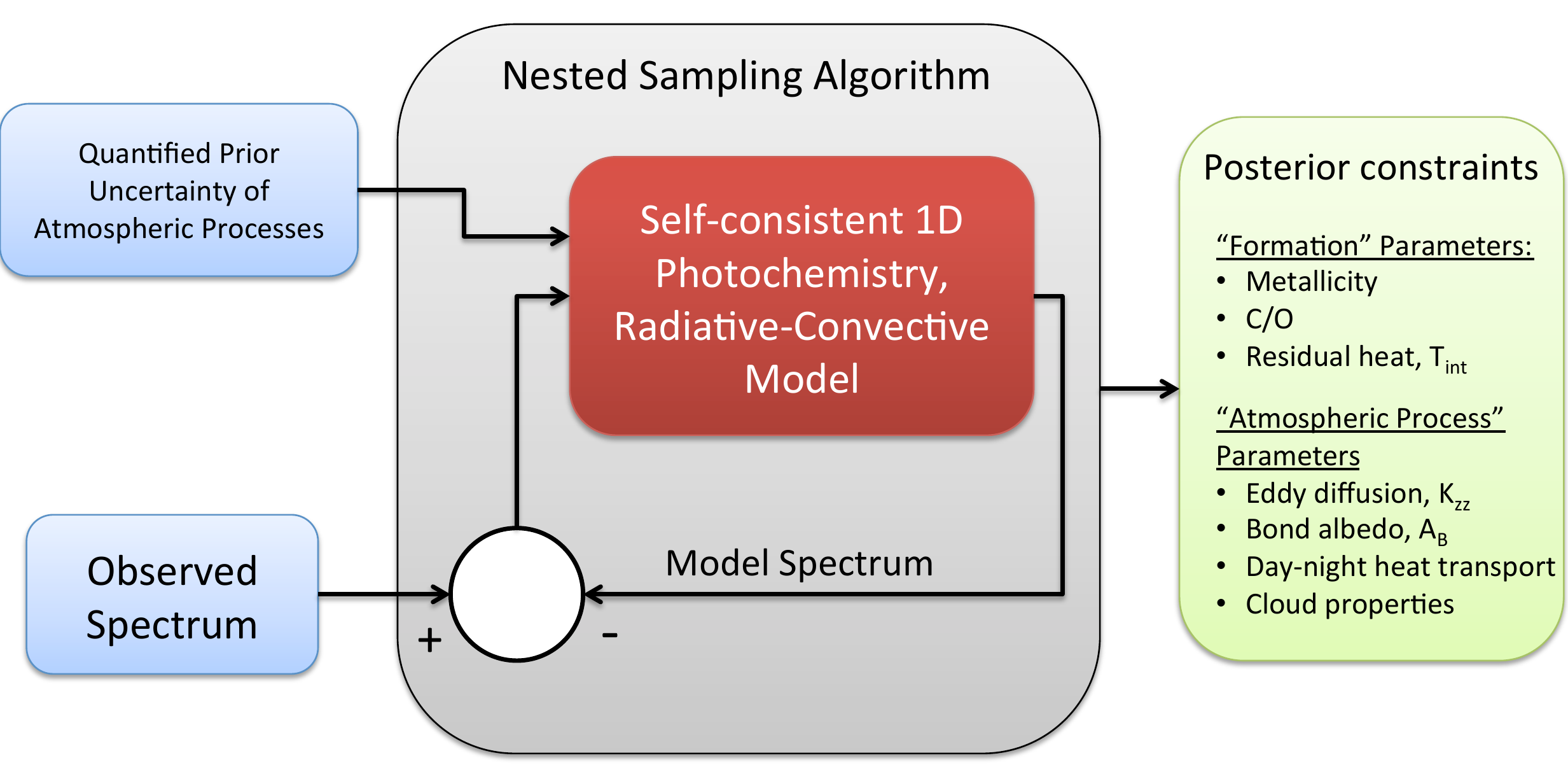}\protect\caption{Schematic view of the SCARLET framework.\label{fig:Scarlet-framework}}
\end{figure}

\section{Methodology: The SCARLET framework}

The main objective of the SCARLET framework is to determine the full
range of self-consistent scenarios for an observed planet by combining
the observational data and our prior knowledge of atmospheric chemistry
and physics in a statistically robust Bayesian analysis. SCARLET consists
of a state-of-the-art self-consistent photochemistry model surrounded
by the nested sampling algorithm for the retrieval. This architecture
enables SCARLET to provide direct insights into the elemental composition
of the deep atmosphere and planetary bulk based on spectral observations
of the upper atmosphere. Compared to previous modeling approaches,
SCARLET combines the statistical robustness and exploratory nature
of traditional atmospheric retrieval methods (Smith et al., 1970;
Rodgers et al. 2000; Madhusudhan \& Seager 2009; Benneke \& Seager
2012) with the self-consistency and ability to provide physical insights
of so-called self-consistent atmospheric forward models (e.g., Burrows
et al. 1997, 2001; Marley et al. 2002; Fortney et al. 2008, 2011;
Moses et al 2013). 

SCARLET is distinct from any previous atmospheric retrieval methods
that describe the current state of the visible atmosphere (molecular
abundance and temperature profiles) by free parameters. The advantage
of the SCARLET approach are as follows: (1) SCARLET provides direct
insight into the elemental composition of the planet's deep atmosphere,
(2) all atmospheric scenarios found by SCARLET are inherently self-consistent
(within the limits of our prior knowledge of atmospheric chemistry
and physics), (3) SCARLET can provide direct constraints on physical
processes such as the strength of vertical mixing in the studied atmospheres.

\subsection{SCARLET Overview}

Figure \ref{fig:Scarlet-framework} illustrates the basic architecture
of the SCARLET framework. The input to the framework are the observed
planetary spectrum and an explicit, quantitative description of our
prior knowledge of atmospheric chemistry and physics. The framework
then repeatedly runs a self-consistent atmospheric chemistry and radiative-convective
model to systemically explore atmospheric scenarios to derive posterior
constraints on the underlying elemental composition of the planet's
gas envelope and the strength and efficiencies of process at play
in the atmosphere.

\subsection{Planet Parameterization\label{sub:Planet-Parameterization}}

The fundamental idea of SCARLET is to divide all unknown properties
of the planet into ``formation history'' parameters and ``atmospheric
process'' parameters. Formation history parameters describe properties
of the planet that are set by the planet's formation and evolution
history, such as its elemental composition and residual accretion
heat. Atmospheric process parameters describe the strengths or efficiencies
of \textit{ongoing} chemical and physical processes in the atmosphere.

\begin{table}[t]
\noindent \centering{}\protect\caption{``Formation'' parameters (top) and ``atmospheric process'' parameters
(bottom) and their prior ranges in the scarlet retrieval analysis
\label{tab:SCARLET-versus-regular}}
\begin{tabular}{cc}
\hline 
SCARLET & Prior range\tabularnewline
\hline 
\hline 
Metallicity at 1000~bar  & $0.1\ldots100$\tabularnewline
C/O at 1000~bar & $0.0001\ldots10000$\tabularnewline
Internal heat  & $0\ldots200\,\mathrm{K}$\tabularnewline
\hline 
Eddy diffusion coefficient, $K_{zz}$  & $10^{7}\ldots10^{11}\,\mathrm{cm^{2}s^{-1}}$ \tabularnewline
Bond albedo & $0\ldots0.9$\tabularnewline
Dayside-nightside heat redistribution factor & $0.25\ldots0.5$\tabularnewline
Cloud properties & see Section \ref{sub:Cloud-Module}\tabularnewline
\hline 
\end{tabular}
\end{table}

\subsubsection{``Formation History'' Parameters}

One of the main goals of studying exoplanets is understanding the
formation and evolution history of planets. Planet properties that
provide hints on the planet's formation histories are the bulk elemental
composition of the planet's gas envelope and the residual interior
accretion heat. While spectroscopic observations of exoplanets cannot
probe these properties directly, SCARLET models the chemistry, radiative
transfer, and mass and heat exchange between the observable optically-thin
atmosphere and the planet deep atmosphere in order to derive conclusion
on the formation history parameters based on spectroscopic observations
of the planet.

\paragraph{Elemental Composition}

In this work, I parameterize the elemental composition of the deep
atmosphere by the metallicity (M) and carbon-to-oxygen ratio (C/O)
at the 1000~bar level. The metallicity is defined as 

\begin{equation}
\mathrm{M=\frac{[X/H]}{[X/H]_{\mathrm{Solar}}}},
\end{equation}

where H is the abundance of hydrogen by number and X is abundance
of any element heavier than helium. A metallicity of 10 times solar
($\mathrm{M}=10$) and the fiducial $\mathrm{C/O=\left(C/O\right)_{Solar}}=0.54$
means that all elements heavier than Helium are ten times more abundant
than they are in the Sun. A log-uniform prior is assigned for metallicities
between 0.1 and 100 (Table \ref{tab:SCARLET-versus-regular}). The
helium to hydrogen ratio is assumed to be equal to the solar value.

The C/O parameter additionally allows for variations in relative abundances
of carbon and oxygen, while keeping the sum of the mixing ratios of
carbon and oxygen constant. C/O ratios between $10^{-5}$ (100,000
times more C) and $10^{5}$ (100,000 times more O) are explored. A
custom-made stretching is applied to ensure an effective exploration
of the parameter space between $\mathrm{C/O=0.1}$ and $\mathrm{C/O=10}$,
where most of the changes in chemistry take place (see y-axis in Figure
\ref{fig:Reduced--of}). A uniform prior probability is applied in
the stretched parameter space, correctly reflecting our prior understanding
that scenarios with extreme C/O ratios are less likely than C/O ratios
between 0.1 and 10.

\paragraph{Interior heat}

The interior heat is described by the effective internal temperature,

\begin{equation}
T_{int}=\left(\frac{F_{Total}}{4\pi\sigma R_{P}^{2}}\right)^{1/4},
\end{equation}

where $F_{Total}$ is the total cooling power of the planet interior,
$4\pi R_{P}^{2}$ is the planet's surface area at the top of the atmosphere,
and $\sigma$ is the Stefan\textendash Boltzmann constant. A uniform
prior on the log scale is between 0~K and 200~K.

\subsubsection{``Atmospheric Process'' Parameters\label{sub:Atmospheric-Process-Parameters}}

The atmospheric process parameters describe chemical and physical
processes in the atmosphere that are weakly understood a priori,
but highly relevant for the interpretation of the observed spectrum.
In this work, the atmospheric process parameters include the eddy
diffusion coefficient ($K_{zz}$), the properties of clouds, the dayside-nightside
heat transport. Atmospheric process parameters are essential for constraining
the desired elemental composition because they provide the connection
between the desired elemental compositions and the observable near-infrared
spectrum of the studied hot Jupiters and enable me to account for
our uncertain prior knowledge of atmospheric processes.

\paragraph{Eddy diffusion}

The eddy diffusion coefficient ($K_{zz}$) is an essential free parameter
because the vertical transport of atmospheric constituents in hot
Jupiter atmosphere is weakly understood and it sensitively affects
how closely the observable upper atmosphere of the planet resembles
the composition in the deep atmosphere. UV photons from the star
break up molecular bonds in the upper atmosphere, thereby significantly
reducing the molecular abundances of species such as methane and water
vapor in the upper atmosphere. The level to which their abundances
are reduced depends sensitively on the vertical transport of atmospheric
constituents that continuously replenish these abundances. Deriving
constraints on the bulk elemental composition of the gas envelope
from molecular absorption formed in the upper atmosphere, therefore,
requires me to account for the uncertainty introduced by the unknown
eddy diffusion coefficient. For hot Jupiters, eddy diffusion coefficients
are uncertain over several orders of magnitude for hot Jupiters ($K_{zz}=10^{7}-10^{10}\,\mathrm{\mathrm{cm^{2}s^{-1}}}$).

\paragraph{Heat redistribution}

The dayside-nightside heat redistribution is parameterized through
a factor, $f$, that relates the irradiation of the 1D model atmosphere
to the stellar flux at the planet's orbital distance from its host
star. The extreme limits are full heat redistribution, i.e. the incoming
radiation is uniformly distributed across the planet, and no dayside-nightside
heat transport, i.e. all the radiation is deposited on the planet's
dayside. The heat redistribution factor ranges from $f=0.25$ for
efficient heat redistribution to $f=0.5$ for no dayside-nightside
heat transport.

\paragraph{Clouds}

Two different cloud modeling approaches are considered in this work.
The ``parameterized particle size and cloud profile'' model captures
the wide range of cloud effects in exoplanet transmission spectra
through four parameters describing the effective grain size ($r_{\mathrm{eff}}$),
the cloud top pressure ($p_{\mathrm{top}}$), the condensate mole
fraction ($q_{*}$), and a cloud profile shape factor ($H_{c}$).
Alternatively, a simplified, two-parameter ``Gray Clouds + Rayleigh
hazes'' model describes the clouds in the atmospheres by specifying
the cloud top pressure of a gray cloud deck and the additional Rayleigh-like
opacity due to small particles. The details of the both cloud modeling
approaches are discussed in Section \ref{sub:Cloud-Module}.

\subsection{Self-Consistent Photochemistry-Thermochemistry Model}

\begin{figure}[t]
\noindent \centering{}\includegraphics[width=1\columnwidth]{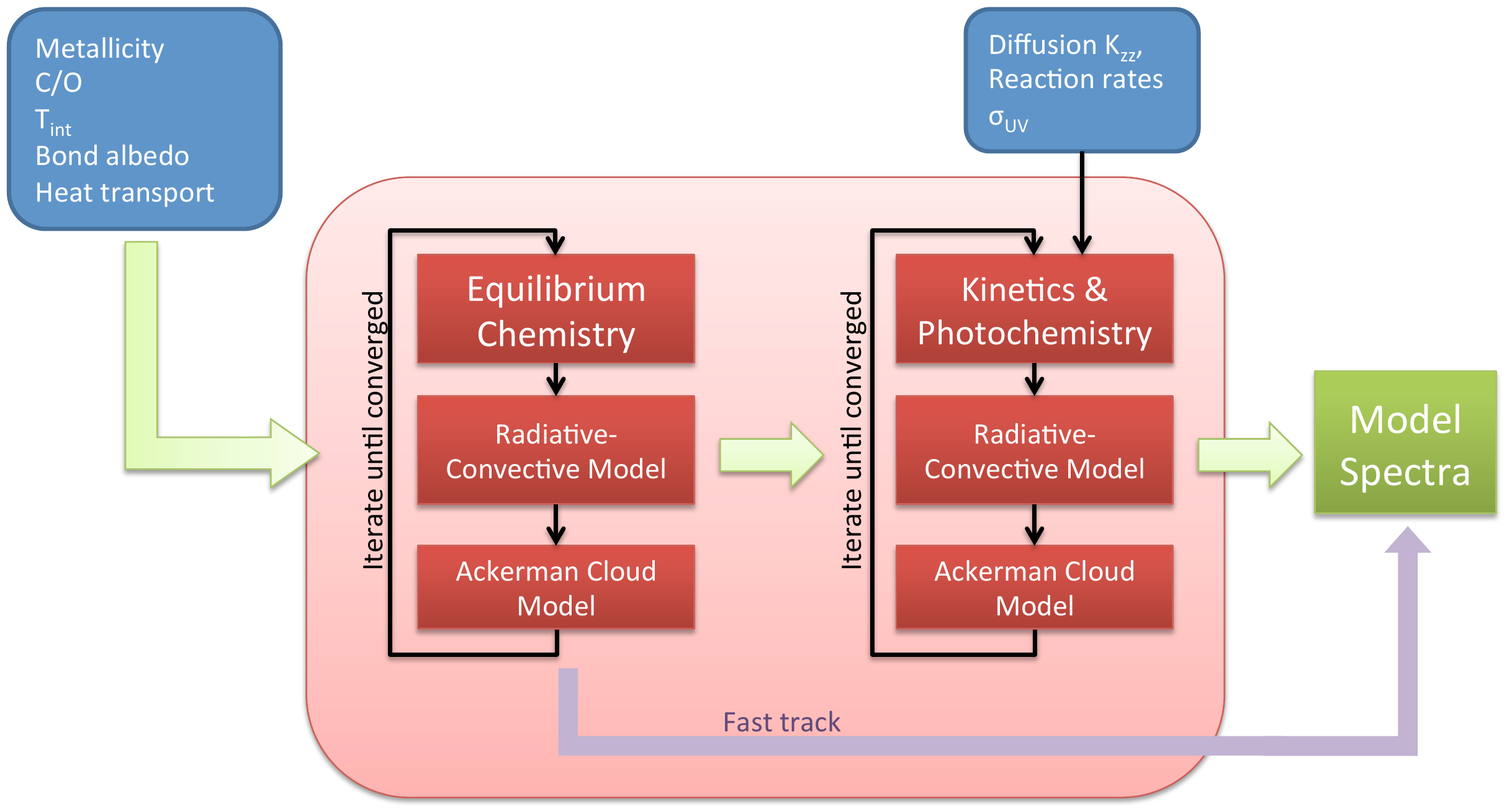}\protect\caption{Flow chart of the self-consistent photochemistry-thermochemistry model
applied in SCARLET. \label{fig:Flow-chart-of}}
\end{figure}

The objective of the photochemistry-thermochemistry model is to compute
self-consistent model atmospheres and synthetic planetary spectra
for any plausible elemental composition, internal heat flux, stellar
irradiation, and atmospheric process parameters. Self-consistent molecular
compositions and temperature structures are obtained by iteratively
solving atmospheric chemistry, radiative-convective heat transport,
and hydrostatic equilibrium. A unique feature of the model is that,
in its most sophisticated form, it self-consistently solves the atmospheric
chemistry via a full thermo- and photochemical kinetics and transport
model and the radiative-convection heat transport via a high-resolution,
line-by-line radiative transfer scheme (Figure \ref{fig:Flow-chart-of}).
This approach enables the derivation of self-consistent molecular
compositions and temperature structures for any plausible elemental
composition. The model complexity can be reduced by approximating
the composition through computationally efficient thermal equilibrium
calculations.

\subsubsection{Chemical Models\label{sub:Chemical-Models}}

Two chemical models with different levels of sophistication are used
in this work. The low-complexity model is a chemical equilibrium model
minimizing Gibb's free energy in each layer of the atmosphere; the
high complexity model is a full thermo- and photochemical kinetics
and transport model. Both models were specifically designed to maximize
computational efficiency to enable thousand of evaluations necessary
for the parameter exploration in SCARLET. The advantage of the chemical
equilibrium model is its computational efficiency. Chemical equilibrium
abundances can be computed $\gtrsim10^{3}$ times faster than performing
a full evaluation of the the thermo- and photochemical kinetics and
transport model. Inferring the elemental abundance of the deep atmosphere
based on the observable upper atmosphere, however, requires capturing
transport-induced quenching in the mid-atmosphere, and photochemistry
in the upper atmosphere, which can only be captured by a full thermo-
and photochemical kinetics and transport model.

\paragraph{Equilibrium chemistry model}

The low-complexity equilibrium chemistry model computes the molecular
compositions from the elemental abundances by minimizing Gibb's free
energy in each layer of the atmosphere (White et al., 1958; \citealp{miller-ricci_atmospheric_2009};
\citealp{benneke_how_2013}). I consider approximately 200 molecular
species from the elements H, He, C, N, O, Ne, Na, Mg, Al, Si, P, S,
Cl, Ar, K, Ca, Ti, Cr, Mn, Fe, and Ni. Rapid computation of $\sim10^{5}$
molecular compositions is accomplished by interpolating the molecular
abundance for each species from a precomputed 4-dimensional equilibrium
grid (30x30x60x30) spanning the dimensions atmospheric metallicity,
C/O ratio, pressure, and temperature.

\paragraph{Kinetics and transport model}

The thermo- and photochemical kinetics and transport model computes
the molecular composition of the atmosphere by solving the coupled
mass-continuity equations as a function of pressure for each molecular
species:

\begin{equation}
\frac{\partial n_{i}}{\partial t}+\frac{\partial\Phi_{i}}{\partial z}=P_{i}-L_{i},\label{eq:continuity}
\end{equation}

where $n_{i}$ is the number density of species $i$ ($\mathrm{m^{-3}}$),
$\Phi_{i}$ is the vertical transport flux of species species $i$
($\mathrm{m^{-2}s^{-1}}$), $P_{i}$ is the chemical production rate
($\mathrm{m^{-3}s^{-1}}$), and $L_{i}$ is the chemical production
rate ($\mathrm{m^{-3}s^{-1}}$), and $z$ is the vertical coordinate
in the atmosphere ($\mathrm{m}$). The chemical production and loss
rates are computed by summing the contributions of 1760 chemical reactions
given the molecular abundance, pressure, temperature in each layer.
The vertical transport flux includes eddy diffusion and molecular
diffusion 
\begin{equation}
\Phi_{i}=-K_{zz}N\frac{\partial f_{i}}{\partial z}-DN\frac{\partial f_{i}}{\partial z}+Dn\left(\frac{1}{H_{0}}+\frac{1}{H_{i}}+\frac{\alpha_{T}}{T}\frac{dT}{dz}\right),\label{eq:-1}
\end{equation}

where $K_{zz}$ is the parameterized eddy diffusion coefficient, $N$
is the total number density, $f_{i}\equiv n_{i}/N$ is the mixing
ratio, $D$ is the molecular diffusion coefficient, $H_{0}$ is the
mean scale height, $H_{i}$ is the scale height of species $i$, and
$\alpha_{T}$ is the thermal diffusion factor.

The kinetics and transport model, in principle, fully captures the
three main chemical processes in planetary atmospheres---thermochemical
equilibrium in the deep atmosphere, transport-induced quenching in
the mid-atmosphere, and photochemistry in the upper atmosphere. A
state-of-the-art reaction list of 1760 chemical reactions for 92 molecular
species formed by the elements H, C, O, and N is adopted from \citet{moses_chemical_2013}.
Thermochemical equilibrium is recovered at high pressure and temperature
because the chemical reaction list includes a matching reverse reaction
for each chemical reaction \citep{visscher_quenching_2011}. 

For the boundary conditions, I assume a zero-flux condition at the
top of the atmosphere boundary and impose the chemical equilibrium
abundances for the specified elemental abundance of the planet's gas
envelope at the lower boundary. The use of a zero-flux condition at
the upper boundary assumes that the atmospheric loss is negligible;
Moses et al. (2011) discuss this assumption in detail and show that
it has no effect on the atmospheric composition in the photosphere.
The lower boundary is set at 1000~bar where temperatures are above
2500~K such that chemical equilibrium mixing ratios will prevail.
The purpose of SCARLET framework is to constrain the elemental abundances
at the lower boundary, i.e. the boundary conditions to the kinetics
and transport model, based on observations of the upper atmosphere.
At the top of the atmosphere, the computational domain reaches to
$10^{-10}\,\mathrm{bar}$; this is sufficiently high to capture the
UV dissociation of all major molecular species. The effect of the
zero-flux condition at the upper boundary neglects atmospheric loss;
throughly discussed in \citet{moses_disequilibrium_2011}, showing
that neglecting atmospheric loss has no effect on the stratospheric
or lower-thermospheric results.

Equations \ref{eq:continuity} and \ref{eq:-1} form a stiff system
of 9200 coupled differential equations for a 100 layer atmosphere,
presenting a substantial computational challenge. The stiffness of
the system results from the vastly different reaction times scales
of the modeled reactions. To efficiently solve the stiff system, I
use a quasi-constant step-size implementation of the Numerical Differentiation
Equation (NDFs). NDFs are found to be more efficient for chemical
kinetics problems as compared with inverse Euler methods generally
used in state-of-the-art atmospheric chemistry codes \citep{allen_vertical_1981,zahnle_atmospheric_2009,hu_photochemistry_2012,moses_chemical_2013}.
Additional acceleration of the computation is accomplished by recomputing
the Jacobian matrix only when it has substantially changed \citep{brown_vode:_1989}.
Despite these improvements, the use of the full the full kinetics
and transport model remains the primary bottleneck for SCARLET retrieval
runs. Evaluation of the kinetics and transport model are initialized
with a converged solution from iterating the chemical equilibrium
and radiative convective model (Figure \ref{fig:Flow-chart-of}).

\subsubsection{Line-by-Line Radiative-Convective Model}

The purpose of the radiative-convective model is to compute temperature-pressure
profiles self-consistent with the stellar insolation and molecular
composition of the atmosphere (Section \ref{sub:Chemical-Models}).
To accomplish this, the radiative-convective model iterates the \textit{T-p}
profile until the radiative downward flux $F^{\downarrow}$ due to
direct star light and infrared reemission is matched by the upward
emission flux in the atmosphere $F^{\uparrow}$. In the process, atmospheric
layers with temperature gradients exceeding the adiabatic lapse rate
$-\frac{dT}{dz}>\Gamma=\frac{g}{C_{p}}$ are declared convective and
adjusted to the adiabatic lapse rate. 

Converging to radiative-convective equilibrium can require hundreds
of radiative transfer calculations of the entire atmospheric column.
In this work, I present a novel numerical technique to speed up repeated
line-by-line radiative transfer calculations by orders of magnitudes
(Appendix A). The new numerical technique enables the efficient computation
of self-consistent T-p profiles for any molecular composition based
on line-by-line equivalent radiative transfer. Line-by-line radiative
transfer has the advantage that it accurately describes the radiative
transport for any atmospheric composition, while analytical solution
for (semi-)gray opacities can deviate significantly from the exact
solution \citep[e.g.,][]{parmentier_non-grey_2014} and the assumption
of correlated opacities across the entire atmospheric column can also
become problematic, in particular for atmospheres with temperature-pressure
profiles near the $\mathrm{CH_{4}/CO}$ or $\mathrm{N_{2}/NH_{3}}$
transition for which the mixing ratios of dominant absorbers can be
strongly altitude-dependent.

\subsubsection{Cloud Models \label{sub:Cloud-Module}}

Clouds play a dominant role in shaping the observable transmission
spectra of exoplanets, yet the formation and chemistry of clouds on
exoplanets are poorly understood. As a result, clouds present a substantial
uncertainty in retrieving the desired atmospheric composition of exoplanets.
Given the lack of a reliable predictive model for clouds on exoplanets,
the approach taken in SCARLET is to describe the clouds properties
using free parameters with the goal of capturing the full range of
plausible cloud properties. Marginalizing over the cloud properties
allows us to determine robust constraints on the planet's gas envelope
composition while accounting for our limited prior understanding of
cloud formation. Two cloud parameterization are applied in this work:

\paragraph{``Parameterized Cloud Profile and Particle Size'' Model}

The ``parameterized particle size and cloud profile'' model captures
a broad range of cloud effects in exoplanet transmission spectra by
exploring different cloud compositions and describing the particle
size and vertical cloud density profile via free parameters in the
fit. Capturing the uppermost structure of the clouds is particularly
important because transmission spectroscopy probes relatively low
pressures as starlight travels on a slant path through the planet's
upper atmosphere \citep{fortney_what_2005}

\begin{figure}[t]
\noindent \begin{centering}
\includegraphics[width=0.4\textwidth]{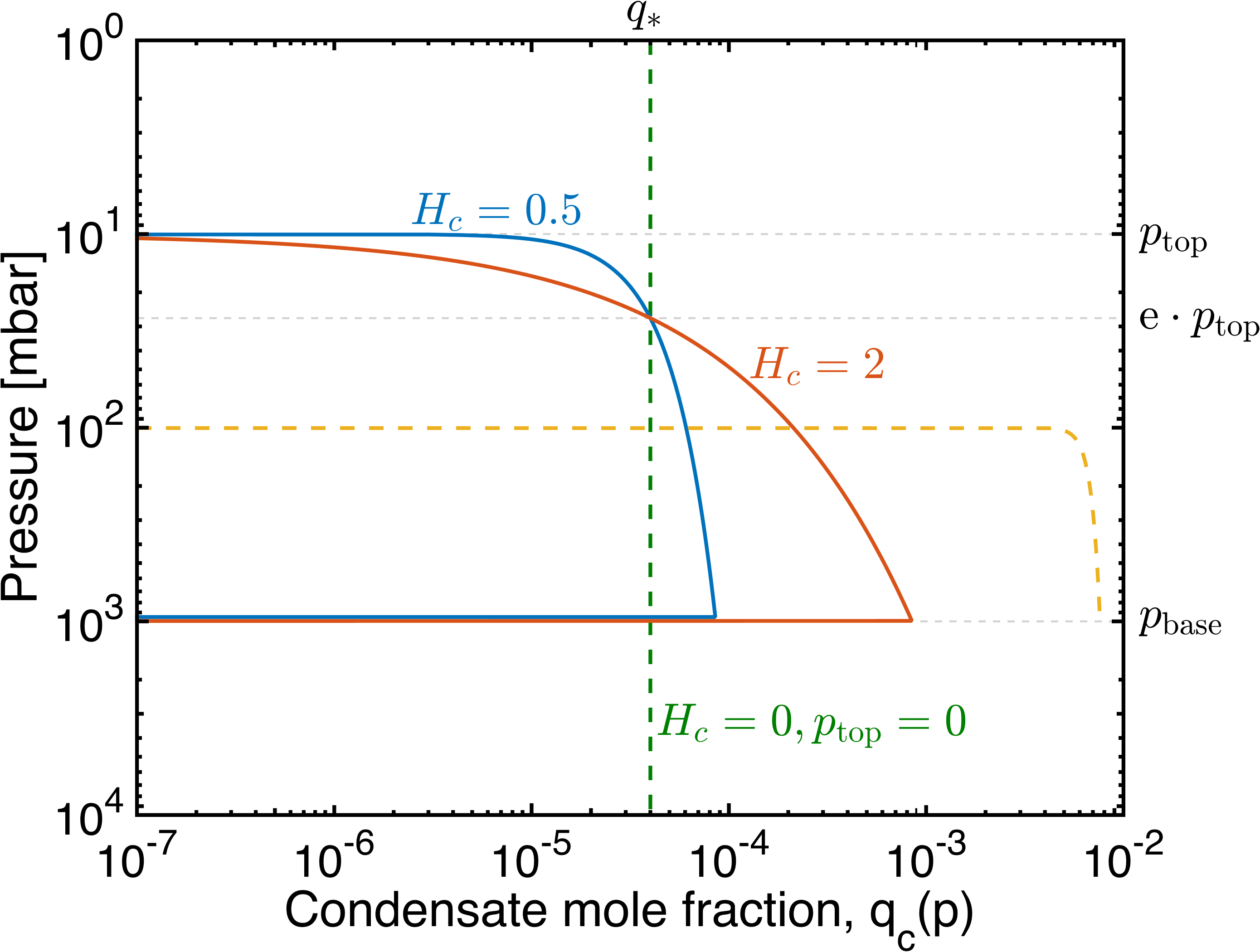}
\par\end{centering}

\noindent \centering{}\protect\caption{Example cloud profiles for the ``parameterized cloud profile and
particle size'' model. The parameterization encompasses a wide range
of cloud profiles including cloud profiles resembling the ones of
typical condensation clouds (red, blue curves --- compare \citet{ackerman_precipitating_2001}),
cloud profiles that assume a uniform condensate mole fraction (green,
dashed line --- compare \citet{etangs_rayleigh_2008}), and sharp
cloud decks (yellow dashed,\citet{benneke_atmospheric_2012}). The
motivation for parameterizing the clouds through free parameters,
rather than modeling them self-consistently, is that our prior knowledge
of the nature and formation of clouds on hot Jupiters is generally
insufficient to reliably capture even the basic trends in the cloud
formation.\label{fig:Parameterized-cloud-density} }
\end{figure}

The approach taken here is to describe the vertical cloud density
profile of the uppermost clouds or hazes by
\begin{equation}
q_{c}\left(p\right)=q_{*}(\log p-\log p_{\mathrm{top}})^{H_{C}}\,\,\,\,\mathrm{for\,}p_{\mathrm{top}}\leq p<p_{\mathrm{base}}\label{eq:cloudprofile}
\end{equation}
where $q_{c}\equiv n_{Cond}/n_{H_{2}}$ is the condensate mole fraction
at pressure $p$, $H_{C}$ is the cloud profile shape factor, $p_{\mathrm{top}}$
is the cloud top pressure, $p_{\mathrm{base}}$ is the pressure at
the cloud base, and $q_{*}$ is the condensate mole fraction one scale
height below the cloud top (Figure \ref{fig:Parameterized-cloud-density}).
The functional form of Equation \ref{eq:cloudprofile} is chosen to
resemble cloud profiles observed and modeled for the solar system
and brown dwarfs.  The particle size distribution is described by
a log-normal distribution 

\begin{equation}
n\left(r\right)=\frac{1}{\sqrt{2\pi}r\log\sigma}e^{-\frac{(\log x-\log r_{\mathrm{eff}})^{2}}{2(\log\sigma)^{2}}},
\end{equation}

where $r_{\mathrm{eff}}$ is the parameterized effective particle
size and the effective variance is set to $\sigma=2$. Altogether,
parameterized particle size and cloud profile describes the uppermost
cloud deck by the four free parameters ($q_{c}$, $p_{c}$, $H_{c}$,
and $p_{top}$). Transmission spectra generally convey little information
about the cloud base, thus $p_{\mathrm{base}}$ is assumed to be high
enough that it has a minimal effect on the transmitted starlight.
Marginalization over the wide range of cloud profiles considered in
the ``parameterized grain size and cloud profile'' model ensures
that retrieved constraints on the atmospheric composition are largely
independent of assumptions on the unknown vertical transport of cloud
particles. 

Particles composed of $\mathrm{MgSiO_{3}}$ , $\mathrm{MgFeSiO_{4}}$,
and $\mathrm{SiC}$ are considered for the hot Jupiters studied in
this work ($T_{\mathrm{eq}}=1410\ldots2010$). The radiative properties
are computed using Mie scattering theory \citep{hansen_light_1974}
and the complex refractive indices taken from \citet{dorschner_steps_1995}.

\paragraph{Gray Clouds + Rayleigh Hazes}

Most published observations of exoplanets provide little information
to observationally constrain the detailed cloud profiles or grain
size. Given sparse data, it can be instructive to investigate the
constraints on cloud top pressure and haze opacity from a low-complexity
``gray cloud deck + Rayleigh hazes'' model. The two-parameter ``gray
cloud deck + Rayleigh hazes ''cloud model simultaneously allows for
the presence of a gray cloud deck as well as ``Rayleigh'' hazes
composed of small particles ($r_{p}\ll\lambda$). 

Gray clouds are modeled as a sharp cutoff to the planet-grazing starlight
below a parameterized pressure level in the atmosphere. A sharp cutoff
approximates the appearance of an upper cloud deck composed on large
particle or a sudden increase cloud density. Rayleigh hazes are assumed
to be composed of small particles ($r_{p}\ll\lambda$). The opacity
of the particles follows $k\propto k_{0.4}\left(0.4\,\mathrm{\mu m}/\lambda\right)^{-4}$,
where $k_{0.4}$ is the extinction coefficient ($\mathrm{cm^{2}/g}$)
at $0.4\,\mathrm{\mu m}$.

\subsubsection{Opacities}

The radiative transfer calculations in this work include opacities
due to molecular absorption, collision-induced broadening from $\mathrm{H_{2}}/\mathrm{H_{2}}$
and $\mathrm{H_{2}/He}$ collisions \citep{borysow_collision-induced_2002},
Rayleigh scattering \citep[see][]{benneke_atmospheric_2012}, and
Mie scattering of cloud and haze particles. Molecular absorption cross
sections are determined directly from the molecular line lists provided
in the high-temperature ExoMol database \citep{tennyson_exomol:_2012}
for $\mathrm{CH_{4}}$, $\mathrm{NH_{3}}$, and $\mathrm{TiO}$, HITEMP
database \citep{rothman_hitemp_2010} for $\mathrm{H_{2}O}$, $\mathrm{CO}$,
and $\mathrm{CO_{2}}$, and the HITRAN database \citep{rothman_hitran_2009}
for $\mathrm{O_{2}}$, $\mathrm{O_{3}}$, $\mathrm{OH}$, $\mathrm{C_{2}H_{2}}$,
$\mathrm{C_{2}H_{4}}$, $\mathrm{C_{2}H_{6}}$, $\mathrm{H_{2}O_{2}}$,
and $\mathrm{HO_{2}}$. Absorption by the alkali metals (Li, Na, K,
Rb, and Cs) is modeled based on the line strengths provided in the
VALD database \citep{piskunov_vald:_1995} and using the $\mathrm{H_{2}}$-broadening
prescription provided in \citet{burrows_calculations_2003}.

To speed up the evaluation of a large number of atmospheric models,
I precompute the wavelength-dependent molecular cross sections for
each of the considered molecular species on a temperature and log-pressure
grid and then interpolate the cross section for the required conditions.
Similarly, the scattering properties of Mie scattering of cloud and
haze particles are precomputed as a function of particle size and
wavelength.

In the upper atmosphere, molecular absorption lines become increasingly
narrow, requiring a very high spectral resolution to exactly capture
the shapes of the thin Doppler-broadened lines \citep{goody_atmospheric_1995}.
Instead of ensuring that each line shape at low pressure is represented
exactly, I choose an appropriate spectral resolution for the line-by-line
simulation by ensuring that the simulated observations are not altered
by more than 1\% of the observational error-bar when the spectral
resolution is doubled or quadrupled.

\begin{figure*}[t]
\noindent \begin{centering}
\hspace{1cm}\includegraphics[width=0.4\textwidth]{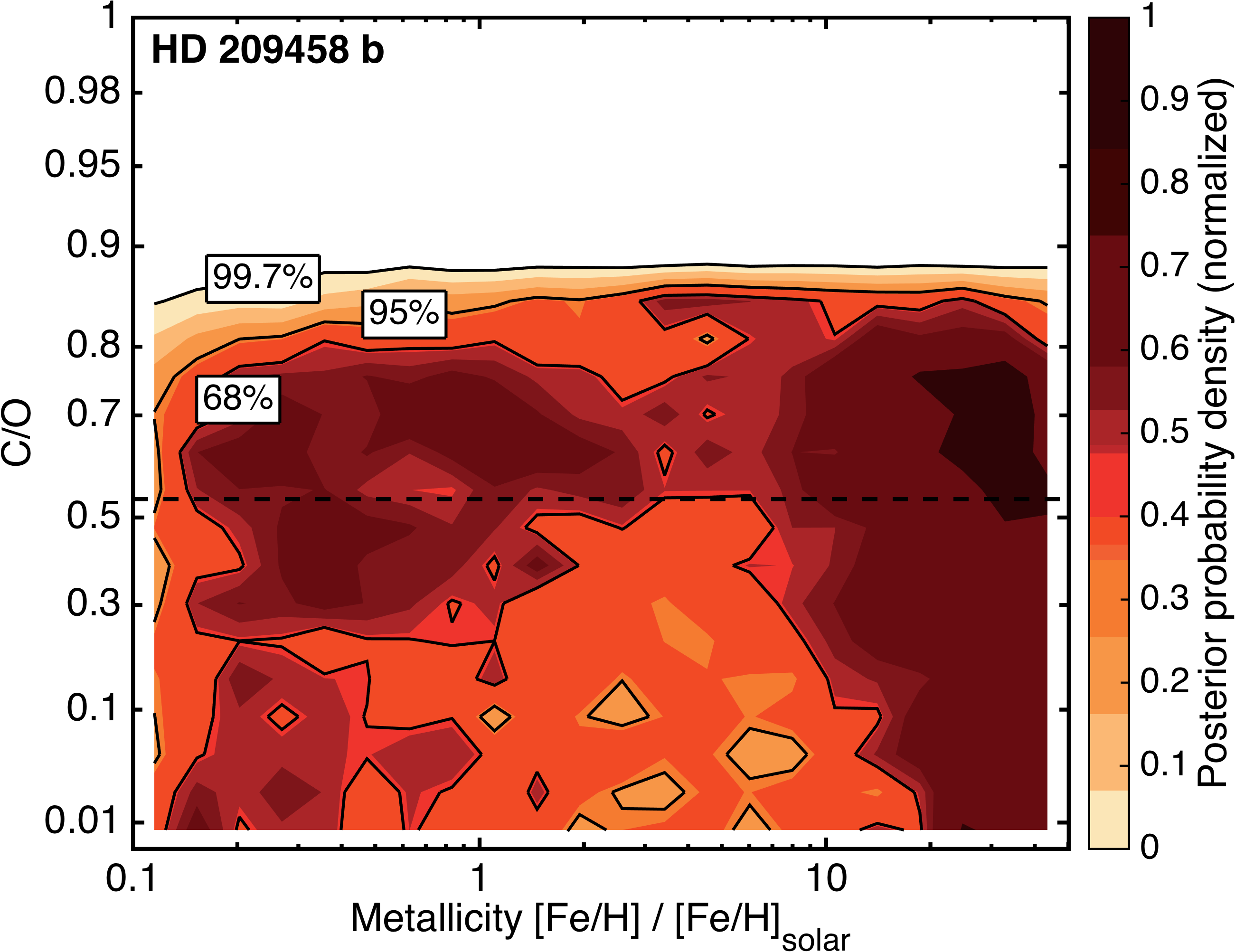}\hfill{}\includegraphics[width=0.4\textwidth]{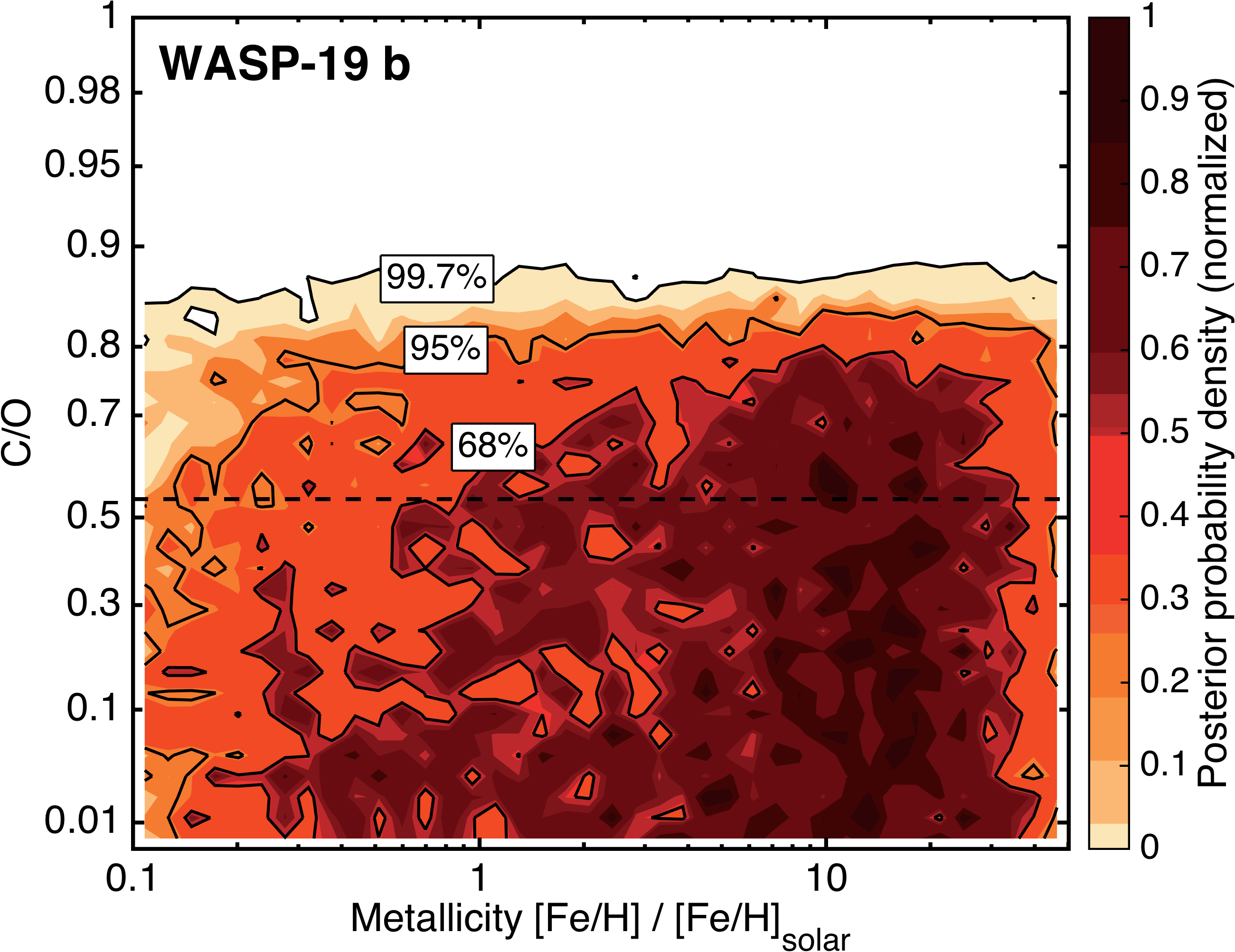}\hspace{1cm}
\par\end{centering}

\noindent \begin{centering}
\hspace{1cm}\includegraphics[width=0.4\textwidth]{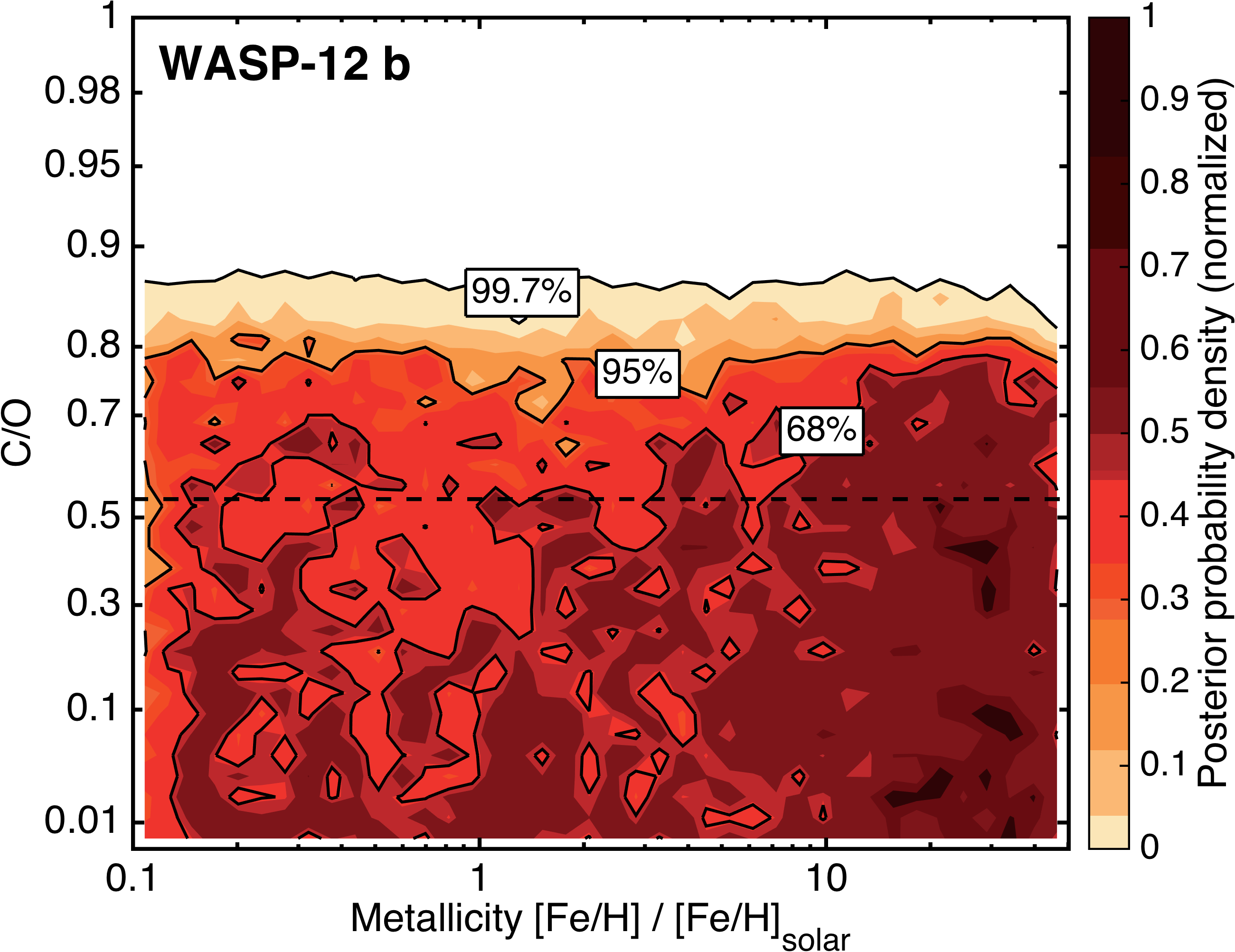}\hfill{}\includegraphics[width=0.4\textwidth]{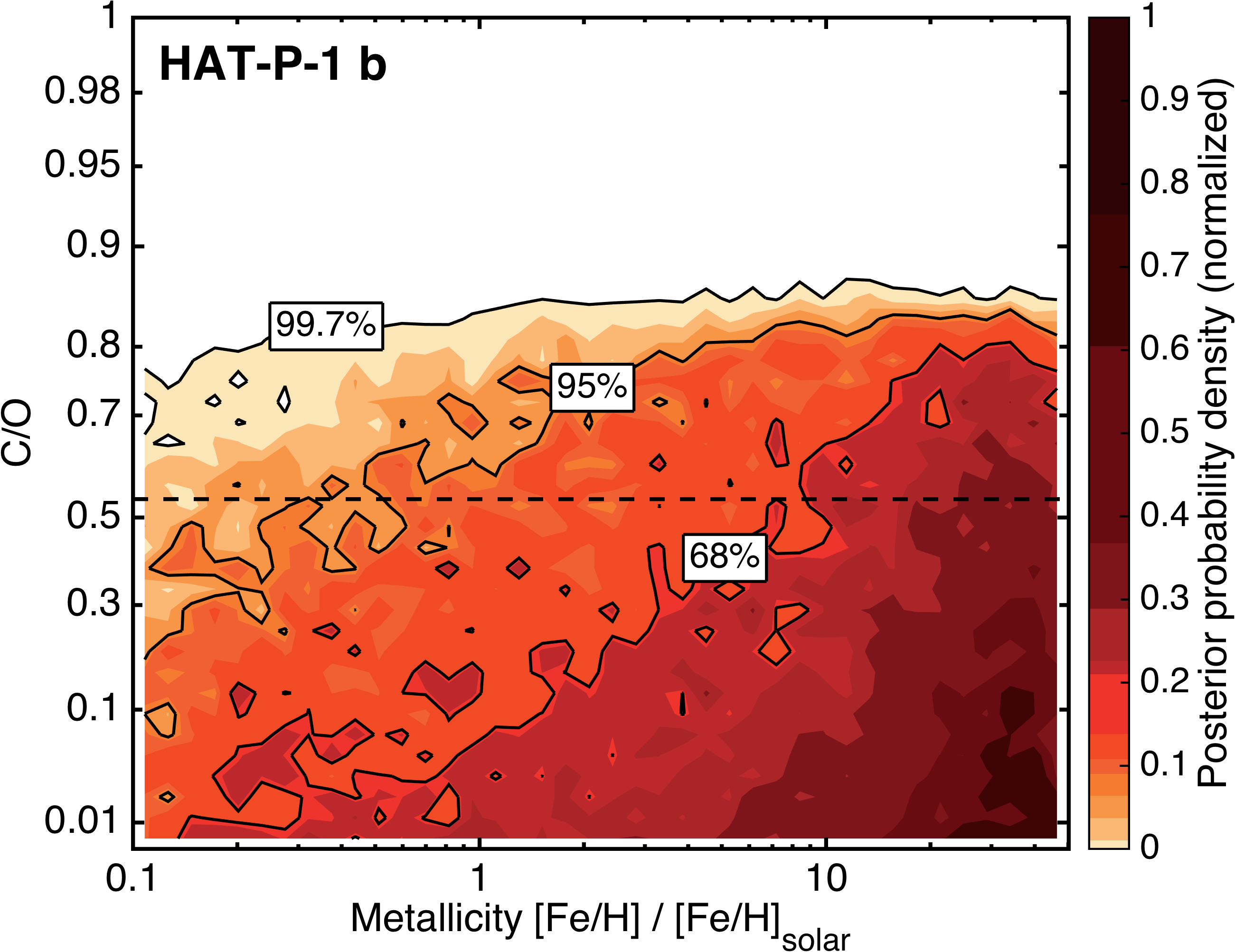}\hspace{1cm}
\par\end{centering}

\noindent \centering{}\protect\caption{Joint constraints on the metallicity and C/O ratio in the deep atmospheres
of HD~209458b, WASP~19b, WASP-12b, and HAT-P-1b. The coloring indicates
the normalized posterior probability marginalized over all remaining
atmosphere and cloud parameters. Black contours mark the 68\% (1$\sigma$),
95\% (2$\sigma$), and 99.7\% (3$\sigma$) Bayesian credible regions.
The \textit{HST WFC3} transmission spectra of all four planets reveal
robustly oxygen-dominated ($\mathrm{C/O<0.9}$) atmospheres. Constraints
on the metallicity, i.e. the overall fraction of species heavier than
He relative to solar, are weak. Results are shown for a cloud model
in which the particle size and vertical cloud density profile are
described by free parameters in the fit (Section \ref{sub:Cloud-Module}).
The posterior distribution is marginalized over the cloud parameters,
the eddy diffusion coefficient, the planetary Bond albedo, the dayside-night
side heat redistribution.\label{fig:Metallicity-vs-CtoO}}
\end{figure*}

\begin{figure*}[t]
\noindent \begin{centering}
\hspace{1cm}\includegraphics[width=0.4\textwidth]{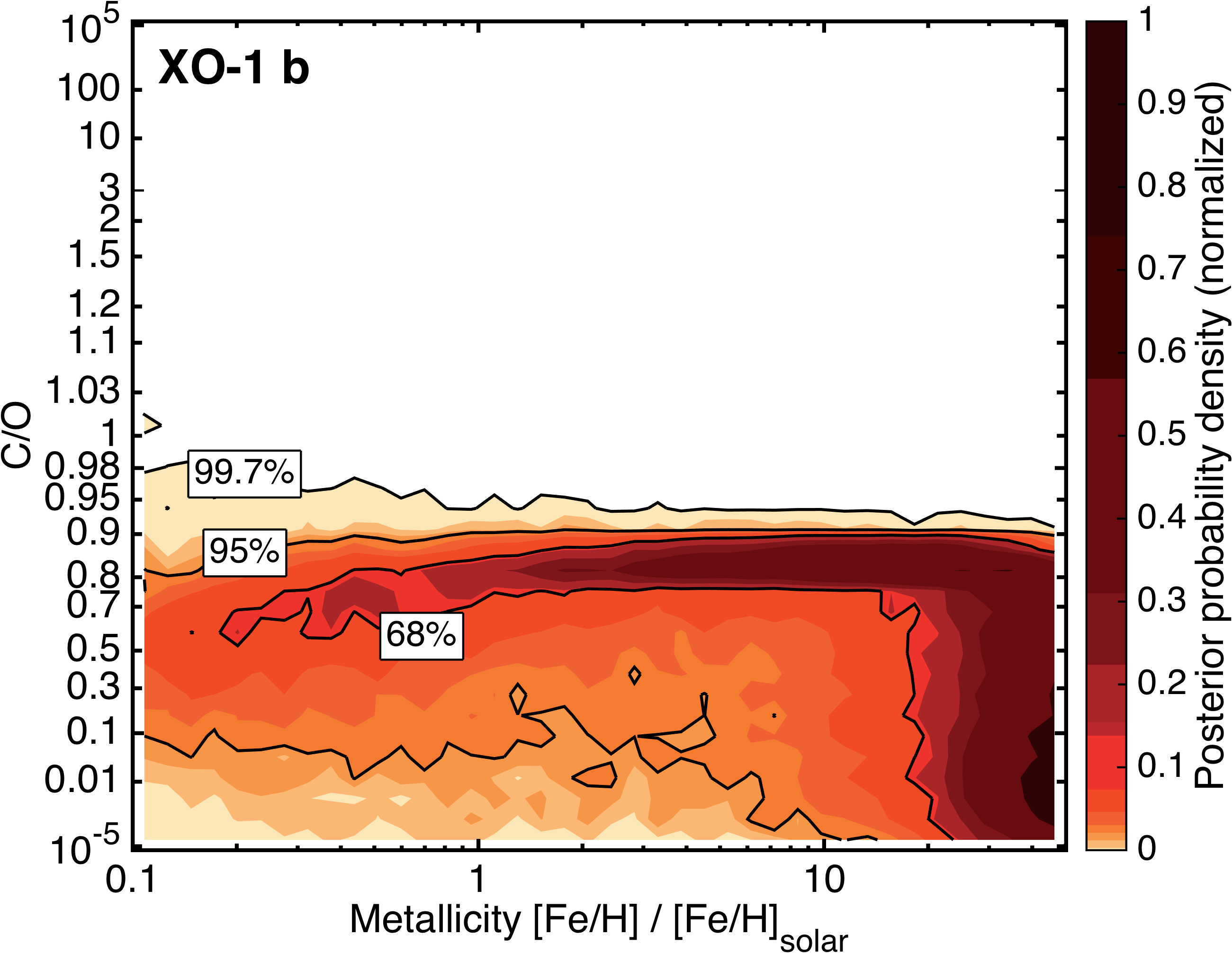}\hfill{}\includegraphics[width=0.4\textwidth]{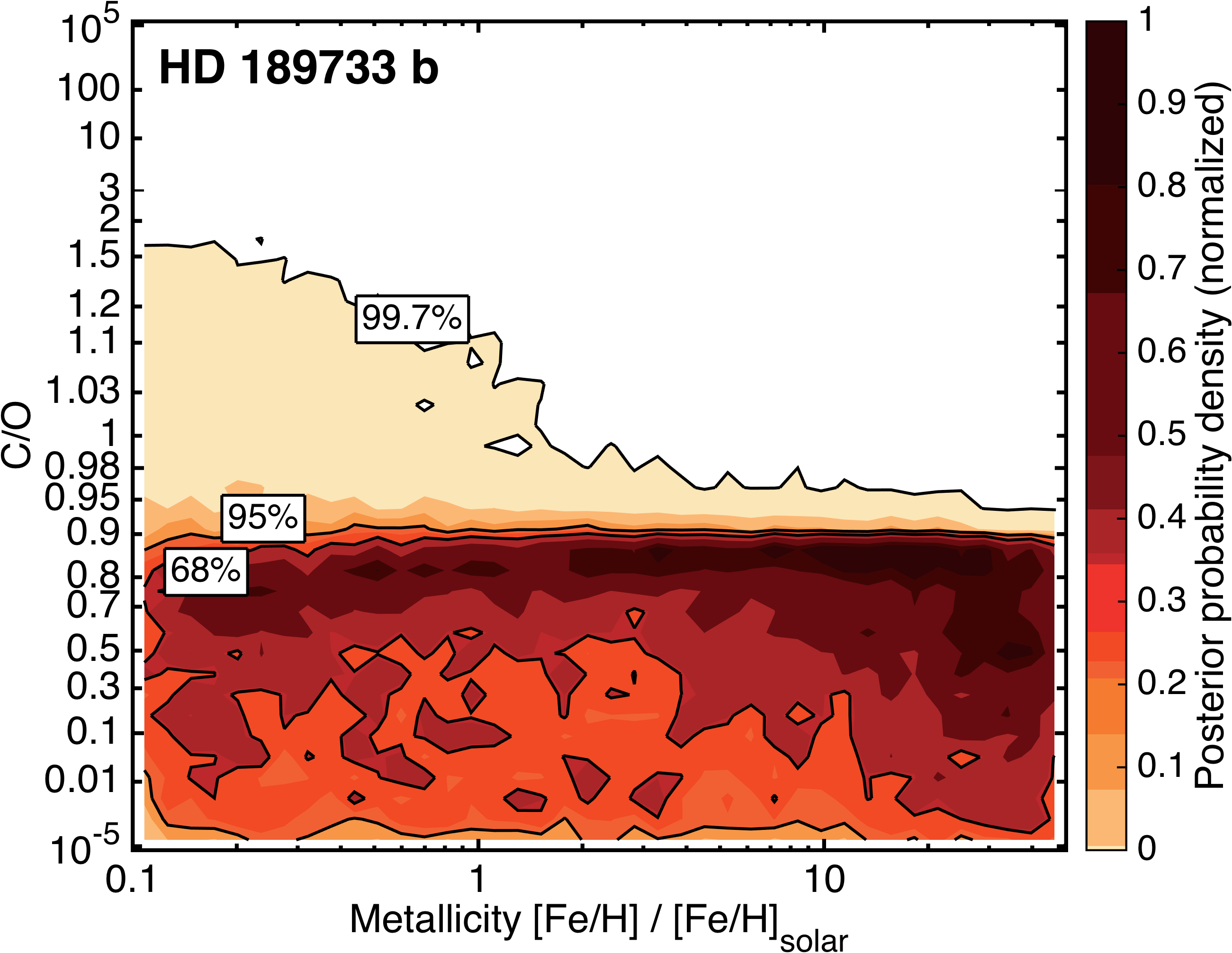}\hspace{1cm}
\par\end{centering}

\noindent \begin{centering}
\hspace{1cm}\includegraphics[width=0.4\textwidth]{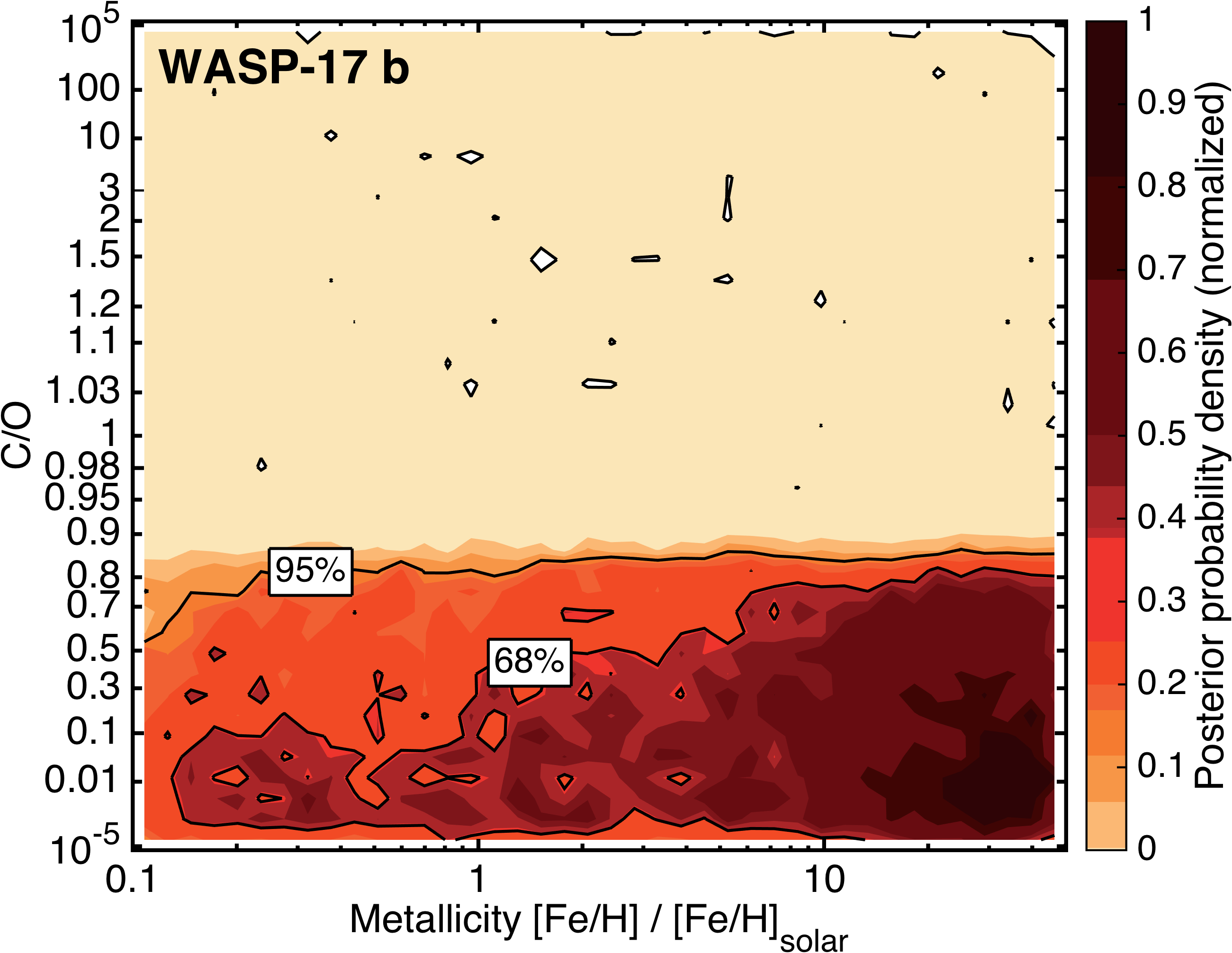}\hfill{}\includegraphics[width=0.4\textwidth]{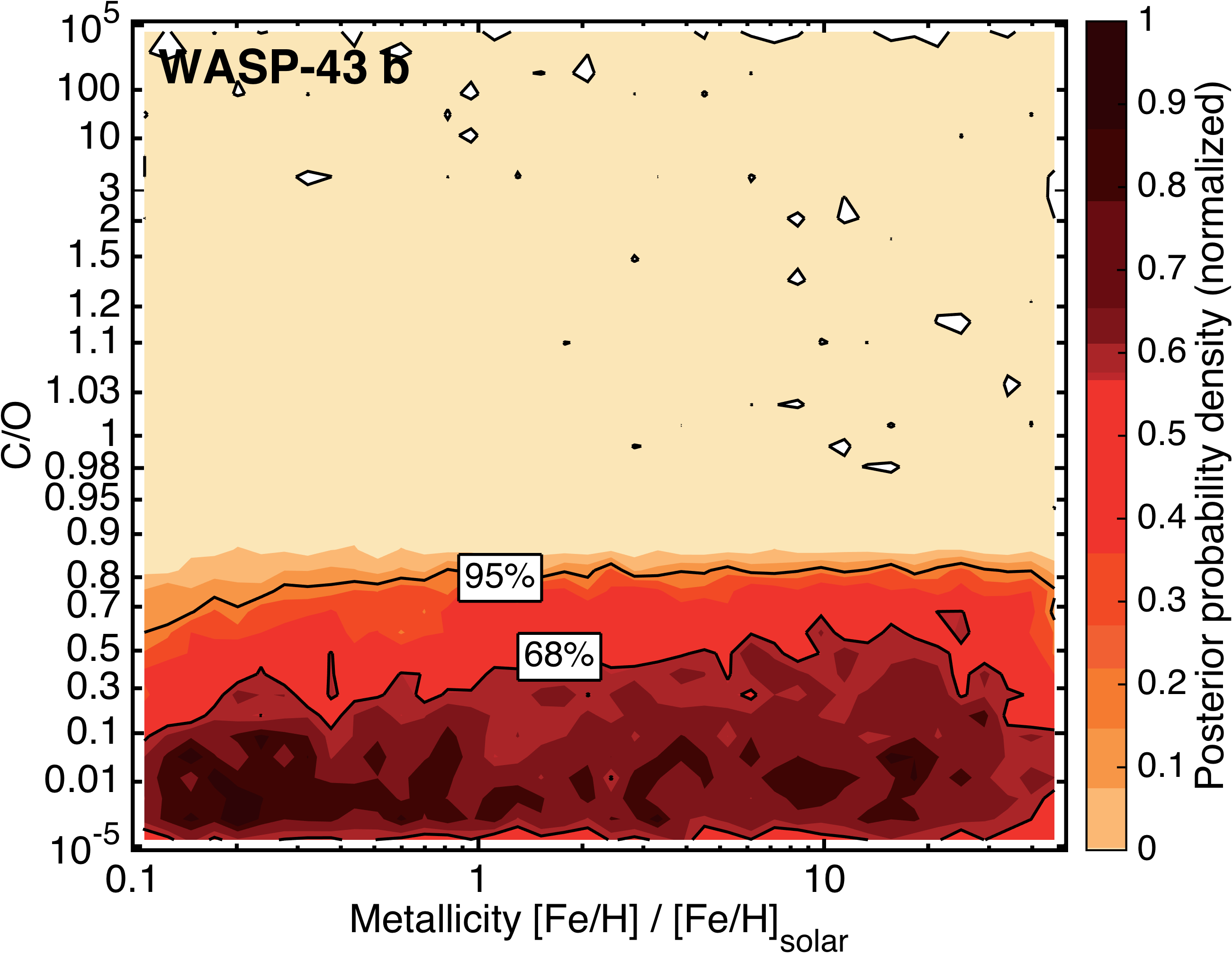}\hspace{1cm}
\par\end{centering}

\noindent \centering{}\protect\caption{Same as Figure \ref{fig:Metallicity-vs-CtoO}, but for the hot Jupiters
XO-1b, HD~189733b, WASP-17b, and WASP-43b. \label{fig:Metallicity-vs-CtoO-1}
The \textit{HST WFC3} transmission spectra of all four planets favor
$\mathrm{C/O<0.9}$ atmosphere at greater 95\% confidence. A low-probability
tail towards high C/O remains for WASP-17b, and WASP-43b containing
a few percent of the posterior probability. To illustrate the tail,
the probability is plotted over the full range from $\mathrm{C/O=10^{-5}}$
to $\mathrm{C/O=10^{+5}}$. \citep{madhusudhan_high_2011,stevenson_deciphering_2014}.}
\end{figure*}

\subsubsection{Transmission Spectrum Model}

Once the chemistry-radiative convective model has converged to a steady-state
solution, I compute model transmission spectra and synthetic instrument
outputs to evaluate the fit between the model and the observations.
The model computes the absorption and scattering of stellar light
by the planetary atmosphere as the rays traverse the day-night terminator
region. Extinction due to molecular absorption, Rayleigh scattering,
and cloud particles absorption and scattering is accounted for. 
In the parameterized cloud profile, the optical properties of finite-size
cloud particles are computed using Mie scattering theory. In the ``gray
cloud deck + Rayleigh haze'' model, the gray cloud deck is modeled
as a sharp cut-off of transmission below the parameterized cloud deck
pressure level. Rayleigh hazes are included as an ad-hoc opacity source
with $\sigma\propto\lambda^{-4}$. Finally, the high-resolution transmission
spectrum is integrated over the instrument response function of the
individual instrument channels for comparison to the astronomical
observations.

\subsection{Nested Sampling for Atmospheric Retrieval\label{sub:Nested-Sampling-for}}

SCARLET employs the multimodal nested sampling algorithm, MultiNest,
to efficiently explore the multidimensional parameter space and compute
the posterior distribution of the formation and atmospheric process
parameters. The mathematical details of the nested sampling algorithm
are described in \citealt{skilling_nested_2004,feroz_multinest:_2009,benneke_how_2013}
and its first application to atmospheric retrieval is discussed in
\citet{benneke_how_2013}. Here, I provide overview of the technique
and its unique capabilities specifically for atmospheric retrieval.

\begin{figure}[t]
\noindent \centering{}\includegraphics[width=1\columnwidth]{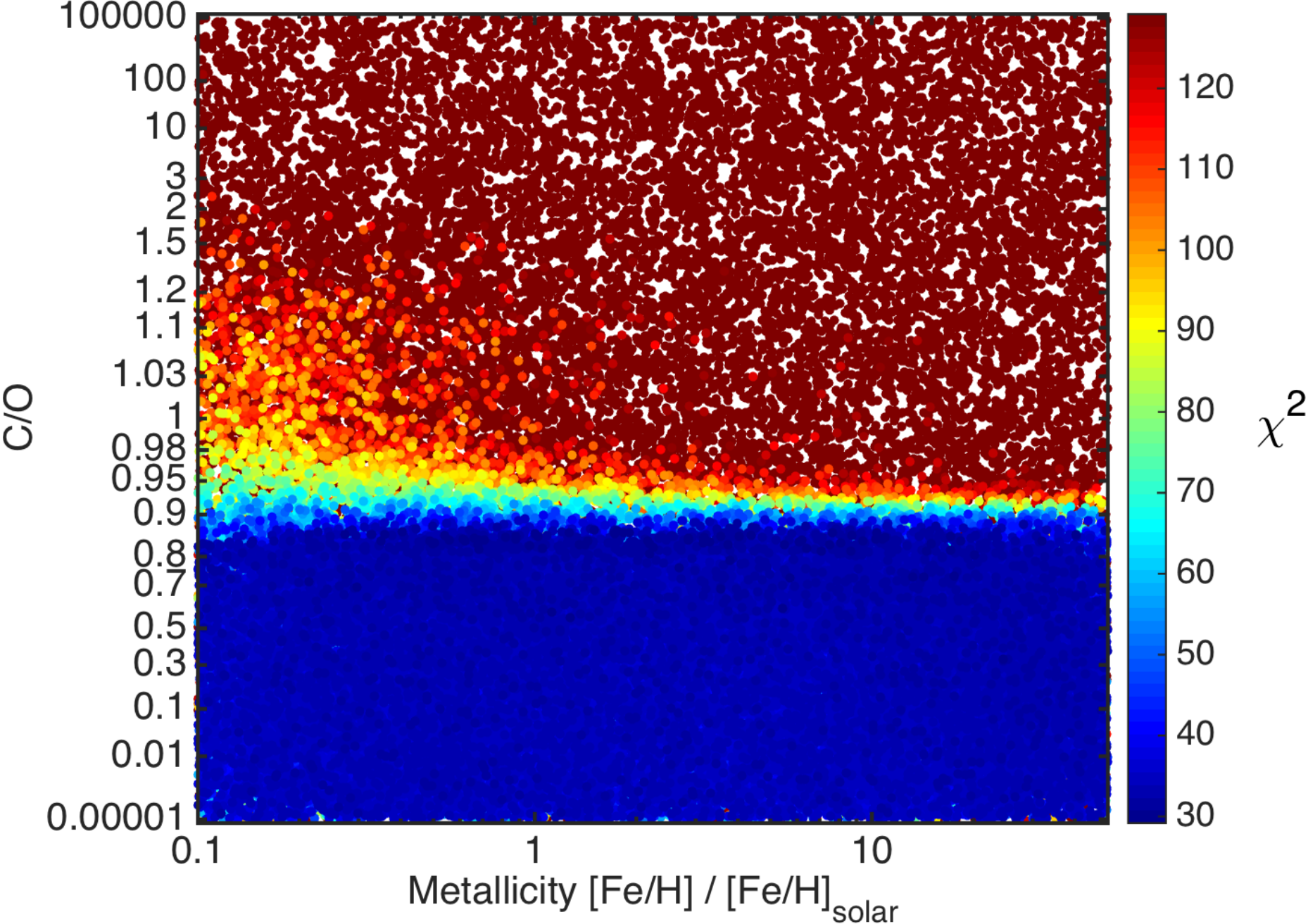}\protect\caption{$\chi^{2}$ of all atmospheric models in the nested sampling results
for HD~209458b. The upper limit $\mathrm{C/O}\lesssim0.9$ is extremely
robust. The nested sampling algorithm explored all corners of the
multi-dimensional parameter space, ranging from $\mathrm{C/O}<10^{-5}$
(100,000x more oxygen) to $\mathrm{C/O}<10^{5}$ (100,000 times more
carbon). All atmospheric scenario with $\mathrm{C/O>1}$ are excluded
at $\Delta\chi^{2}>55.9$, corresponding to a likelihood ratio of
$1.4\cdot10^{12}\,\mathrm{to}\,1$. The fit to the data is excellent
($\chi^{2}/N=1.1$) $\mathrm{C/O}\lesssim0.9$ and drops off sharply
near $\mathrm{C/O=0.9}$ to $\chi^{2}/N\gtrsim4$. \label{fig:Reduced--of}}
\end{figure}

Similar to the widely used Markov Chain Monte Carlo (MCMC), nested
sampling is a Monte Carlo approach to efficiently compute the multidimensional
posterior distribution of model parameters in parameter estimation
problems. The main advantage of nested sampling for atmospheric retrieval
is, however, that nested sampling reliably captures highly non-Gaussian
and multimodal posterior distributions \citep[e.g.][]{feroz_multinest:_2009,benneke_how_2013}.
This is extremely important for atmospheric retrieval because strong
correlations and degeneracies between atmospheric parameters are common
due to the generally sparse data available for exoplanets and the
complex way in which vital information about planet is encoded in
the observable planetary spectrum. Importantly, nested sampling also
provides an excellent overview of the goodness-of-fit all across the
entire parameter space, even in regions far away from the maximum
probability (e.g., Figure \ref{fig:Reduced--of}). This provides an
improved ability to assess which regions of the parameters space can
be excluded at levels of significance higher than the commonly stated
95\% or 99.7\% probability. 

Nested sampling begins by randomly taking a user-specified number
of ``active samples'' ($N\approx100\ldots10000$) from the \textit{entire}
prior parameter space. The active samples are then migrated towards
regions of high likelihood by repeatedly replacing the lowest-likelihood
sample by a new sample with a likelihood higher than the sample to
be rep. In this ``outside-in'' approach, the complete prior parameter
space in traversed providing high confidence that the global minimum
is identified and that extremely correlated or banana-shaped posteriors
are captured correctly. Replaced samples are stored such that a relatively
good description of the probability across the entire parameters space
can be inferred.  In this work, I employ simultaneous ellipsoidal
nested sampling to efficiently find sample with likelihoods higher
than the previously lowest-likelihood sample \citep{feroz_multinest:_2009}.

\section{Results\label{sec:Results}}

\begin{figure*}[t]
\noindent \begin{centering}
\includegraphics[width=0.8\textwidth]{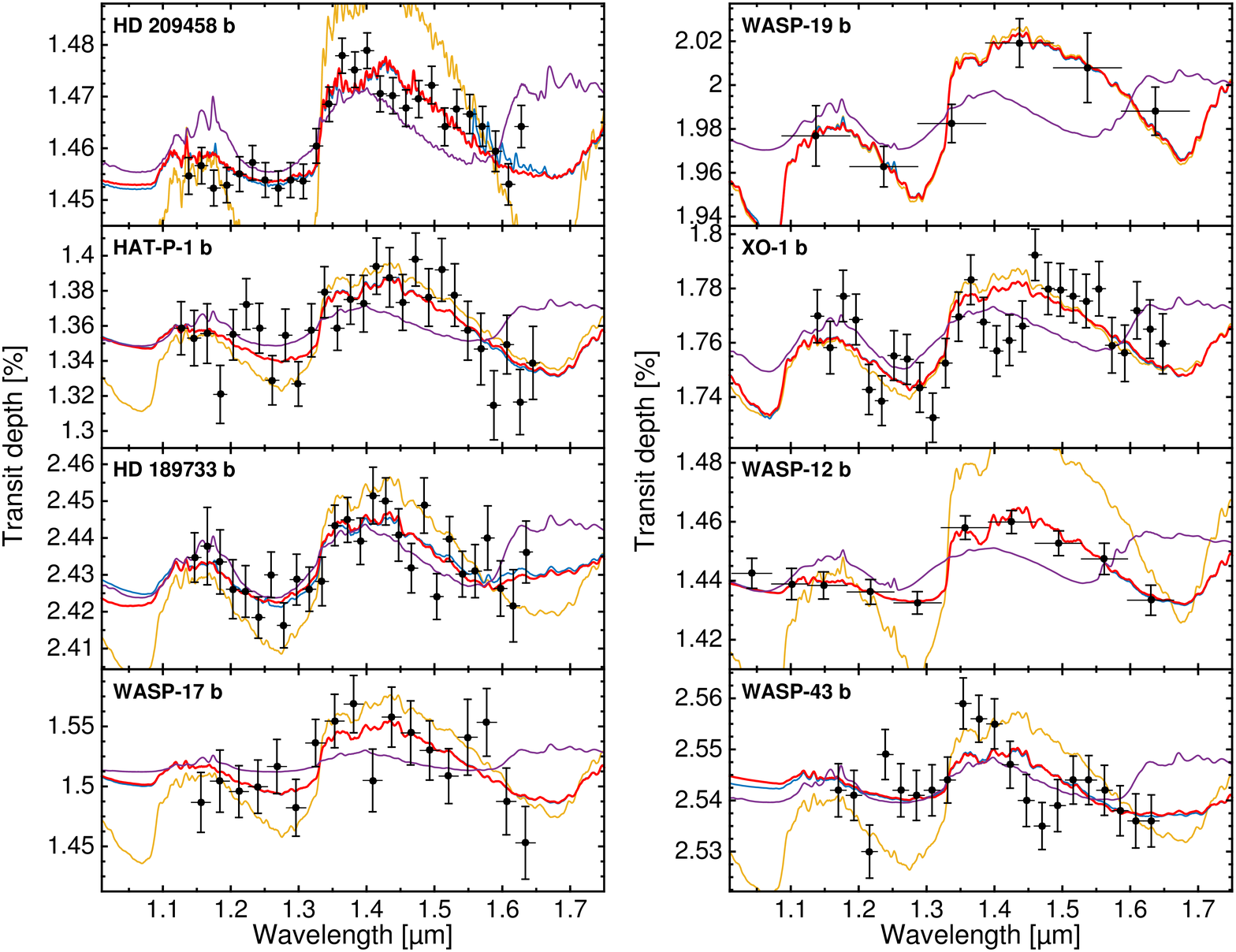}
\par\end{centering}

\noindent \centering{}\protect\caption{Model transmission spectra compared to the published \textit{HST WFC3}
transit depth measurements (black). Solid lines show the transmission
spectra for a fiducial clear solar composition atmosphere (yellow),
the overall best-fitting model (red), the best-fitting solar composition
atmosphere with clouds (blue), and the best fitting C/O>1 model with
clouds (purple). Plotted observations and their $1-\sigma$ uncertainties
are taken from \citet{deming_infrared_2013} for HD~209458b and XO-1b,
\citet{huitson_hst_2013} for WASP-19b, \citet{wakeford_hst_2013}
for HAT-P-1b, \citet{mccullough_water_2014} for HD~189733b, \citet{kreidberg_detection_2015}
for WASP-12b, \citet{mandell_exoplanet_2013} for WASP-17b, and \citet{kreidberg_precise_2014}
for WASP-43b. The model spectra have a vast number of molecular lines;
for clarity the spectrum has been Gaussian smoothed. \label{fig:Model-transmission-spectra}}
\end{figure*}

\begin{figure}[t]
\noindent \centering{}\includegraphics[width=1\columnwidth]{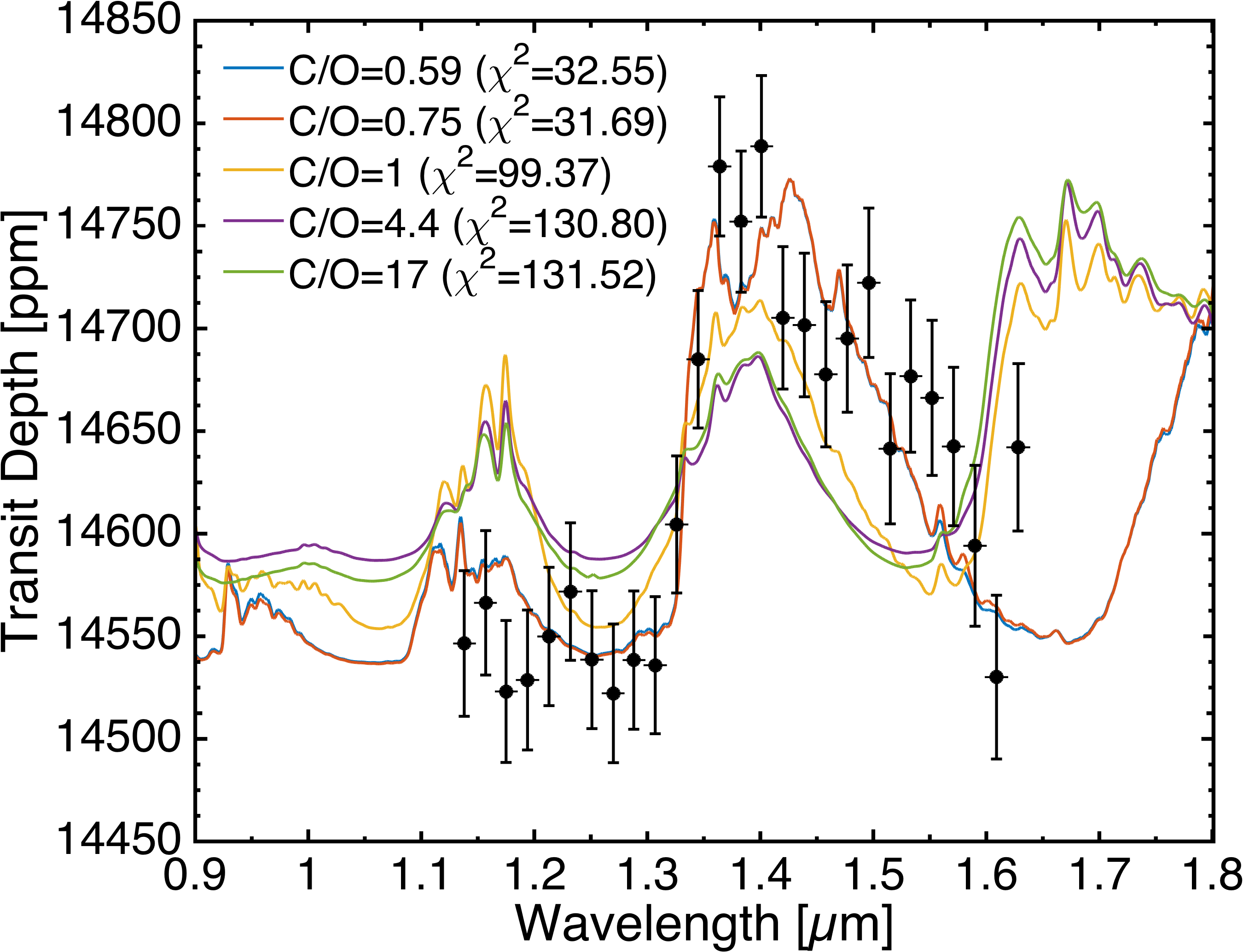}\protect\caption{Effect of the C/O ratio on the \textit{WFC3} transmission spectrum
of HD~209458b. Colored curves show the best-fitting model spectra
at C/O=0.59 (blue), C/O=0.75 (red), C/O=1 (yellow), C/O=4 (purple),
and C/O=17 (green). Best-fitting model are obtained by fixing the
C/O ratio and optimizing all other model parameters including the
cloud properties. Water and methane have strongly overlapping absorption
bands in the WFC3 bandpass \citep{benneke_how_2013}; the increased
transit depth at 1.15 and 1.4~$\mathrm{\mu m}$ is due $\mathrm{H_{2}O}$
absorption for $\mathrm{C/O<1}$ (oxygen-dominated chemistry), but
due to $\mathrm{CH_{4}}$ absorption for $\mathrm{C/O>1}$ (carbon-dominated
chemistry). All best-fitting models contain a thick cloud deck between
2 and 17~mbar to match the relative small amplitude of the observed
absorption features. $\mathrm{C/O<1}$ is in good agreement with the
observations --- clouds in the 1-10~mbar regime mute the weaker 1.15~$\mathrm{\mu m}$
$\mathrm{H_{2}O}$ absorption features, while preserving the stronger
$\mathrm{1.4\,\mu m}$ feature. $\mathrm{C/O>1}$ is in strong disagreement
with the data; however, because the methane absorption bands at 1.15
and 1.4~$\mathrm{\mu m}$ are similarly strong and would inevitably
result in two similarly strong absorption features at 1.15 and 1.4~$\mathrm{\mu m}$,
which is not observed.  \label{fig:Reduced--of-1}}
\end{figure}

\begin{figure}[t]
\noindent \begin{centering}
\includegraphics[width=1\columnwidth]{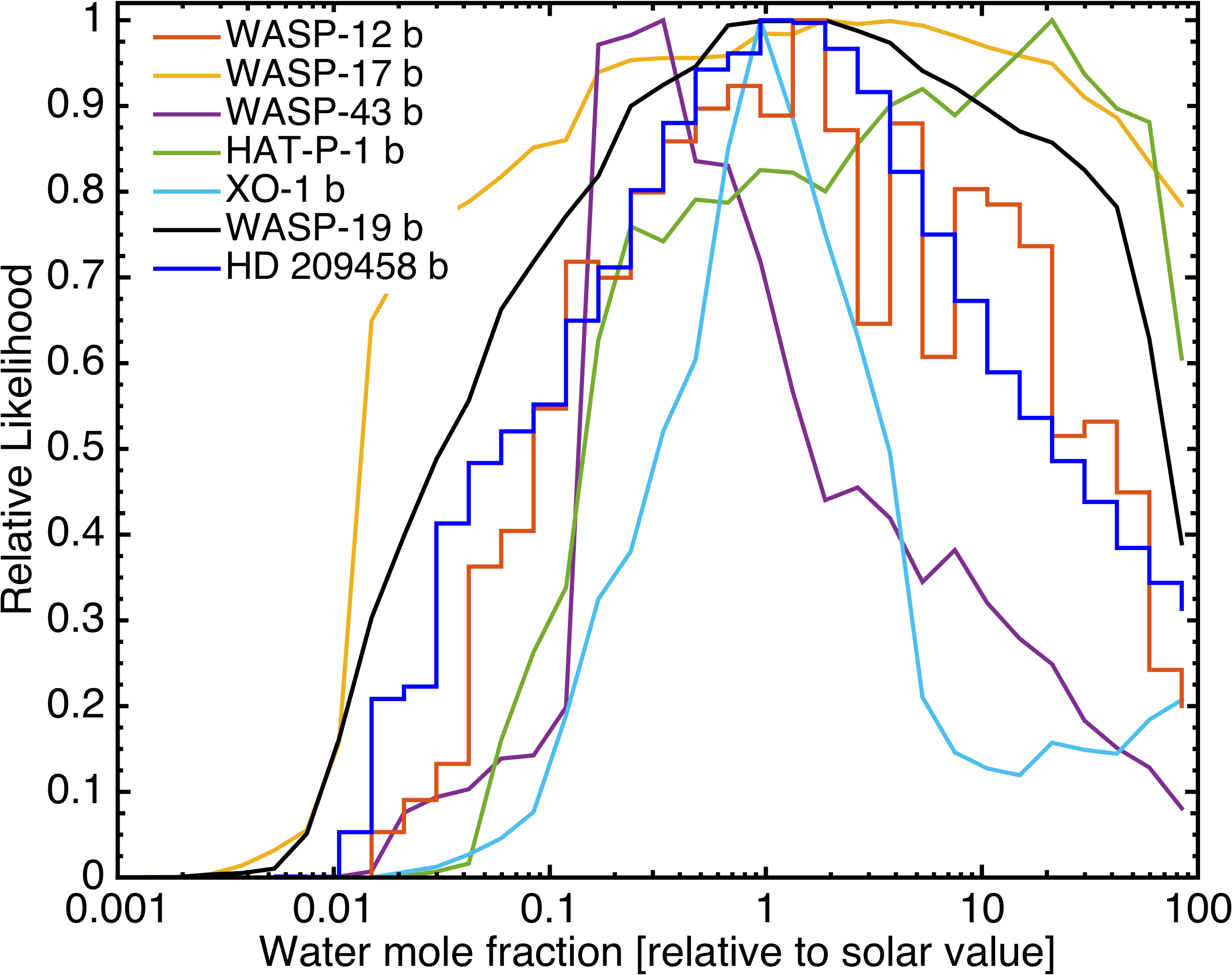}
\par\end{centering}

\noindent \centering{}\protect\caption{Constraints on the water mole fraction for eight hot Jupiters. Colored
curves indicate the relative likelihood of the water mole fraction
at 100~mbar relative to the value expected for solar composition
atmospheres ($\sim$610~ppm). The water abundances of all eight planets
are in agreement with solar composition. A relatively wide range of
water abundances is in agreement with the observations due to strong
correlation between the water mole fraction and the cloud top pressure
(see also Figure \ref{fig:Water-vs-Clouds}). Accounting for clouds
in the atmospheric retrieval model, I do not confirm the inference
of sub-solar water abundances for HD~209458b and HD~189733b \citep{madhusudhan_h2o_2014}.
\label{fig:Water-abundance}}
\end{figure}

\begin{table*}[t]
\noindent \centering{}\protect\caption{Summary of planet and stellar temperatures, goodness-of-fit, and C/O
constraints.\label{tab:SCARLET-versus-regular-1-1}}
\begin{tabular}{ccccccccc}
\hline 
Planet & HD 209458b & WASP-19b & WASP-12b & HAT-P-1b & XO-1b & HD 189733b & WASP-17b & WASP-43b\tabularnewline
\hline 
\hline 
$T_{\mathrm{Eq}}$ {[}K{]} for $A_{B}=1$ & 1409 & 2013 & 2517 & 1271 & 1176 & 1169 & 1508 & 1340\tabularnewline
$T_{\ast,\mathrm{eff}}$ $\left[\mathrm{K}\right]$  & 6065 & 5500 & 6300 & 5975 & 5750 & 5040 & 6550 & 4400\tabularnewline
\hline 
$\chi_{\mathrm{min}}^{2}$ & 30.8 & 1.4 & 4.0 & 26.8 & 35.5 & 17.5 & 14.4 & 24.8\tabularnewline
$\chi_{\mathrm{min}}^{2}/N$ & 1.07 & 0.37 & 0.58 & 0.98 & 1.13 & 0.81 & 0.89 & 1.09\tabularnewline
\hline 
C/O (95\% limit) & < 0.86 & < 0.82 & < 0.80 & < 0.77 & < 0.87 & < 0.92 & < 0.89 & < 0.87\tabularnewline
C/O (99.7\% limit) & < 0.88 & < 0.85 & < 0.87 & < 0.84 & < 0.98 & < 1.20 & --- & ---\tabularnewline
$\left(\Delta\chi^{2}\right)_{\mathrm{min}}$ for C/O>1 & 55.9 & 13.3 & 36.0 & 22.9 & 12.6 & 7.1 & 10.0 & 6.9\tabularnewline
$\mathcal{L}_{\mathrm{max}}/\mathcal{L}_{\mathrm{C/O>1}}$  & $1.4\cdot10^{12}\,:\,1$ & 788 : 1 & $7.0\cdot10^{7}\,:\,1$ & 99,720 : 1 & 553 : 1 & 35 : 1 & 153 : 1 & 32 : 1\tabularnewline
\hline 
\end{tabular}
\end{table*}

In this section, I present constraints on the atmospheric composition
and cloud properties in eight hot Jupiters (HD~209458b, WASP-19b,
HAT-P-1b, and XO-1b, HD~189733b, WASP-12b, WASP-17b, and WASP-43b)
based on their measured transmission spectra. To capture the uncertainty
in the atmosphere composition introduced by the broad variety of plausible
cloud properties, I present constraints on the composition based on
the ``parameterized particle size and cloud profile'' model. Cloud-introduced
uncertainties on the composition are fully accounted for by marginalizing
over the wide range of cloud parameters in this model (Section \ref{sub:C/O-and-Metallicity}-\ref{sub:Water-Abundance}).
The compositional constraints presented in this section solely rely
on the \textit{HST} \textit{WFC3} observations, and do not depend
on comparisons of observations taken at different times and different
instruments. In Section \ref{sub:Clouds-and-Hazes}, I then present
constraints on the cloud properties using both the ``parameterized
particle size and cloud profile model'' as well as a simplified ``gray
cloud deck + Rayleigh haze model''. \textit{HST STIS} and \textit{Spitzer}
observations are included to supplement to conclusions on cloud properties.

\subsection{C/O and Metallicity\label{sub:C/O-and-Metallicity}}

The main result of this study is that all eight hot Jupiters with
detectable near-infrared absorption features show a strict upper limit
on the C/O ratio at approximate 0.9. Carbon-rich atmospheric compositions
(C/O>1) are firmly ruled out for HD~209458b, WASP-19b, WASP-12b,
HAT-P-1b, and XO-1b --- virtually no posterior probability exists
for atmospheric compositions with C/O>0.9 (Figure \ref{fig:Metallicity-vs-CtoO}).
The remaining two hot Jupiters (HD~189733b, WASP-17b, and WASP-43b)
show similar posterior distributions; however, a low-probability tail
(<5\% probability) remains towards high C/O given currently available
data (Figure \ref{fig:Metallicity-vs-CtoO}). Meanwhile, the available
observations provide virtually no constraint on the atmospheric metallicity
for any of the planets.

The uppers limit of $\mathrm{C/O}<0.9$ is a robust finding. The nested
sampling algorithm explores every corner of the multi-dimensional
parameter space, ranging from $\mathrm{C/O}<10^{-5}$ ($10^{5}$ times
more oxygen) to $\mathrm{C/O}<10^{5}$ ($10^{5}$ times more carbon),
and there is no atmospheric scenario with C/O>0.9 that results in
a good fit to the data. As an example, the fit to the data degrades
sharply from $\chi_{\mathrm{}}^{2}/N=1.1$ to worse than $\chi_{\mathrm{}}^{2}/N=4$
for C to O ratios above 0.9. (Figure \ref{fig:Reduced--of}).

The chemical explanation for the strict upper limit on the C/O ratio
is that the observed water absorption feature at $1.4\,\mathrm{\mu m}$
can only exist if substantially more oxygen is present in the gas
envelope than carbon (Figure \ref{fig:Model-transmission-spectra}).
As the C/O ratio approaches unity, almost all of the oxygen becomes
trapped in CO molecules leaving no oxygen left to form $\mathrm{H_{2}O}$.
The upper limit on the C/O ratio is largely independent of the metallicity
because the atmospheric chemistry at the high temperatures encountered
in hot Jupiters sharply changes between $\mathrm{C/O=0.8}$ and $\mathrm{C/O=1}$.

Methane and water have strongly overlapping absorption bands in the
WFC3 bandpass at at 1.15 and 1.4~$\mathrm{\mu m}$ (Figure \ref{fig:Reduced--of-1}).
Nonetheless, I can unambiguously distinguish between $\mathrm{H_{2}O}$
absorption (oxygen-dominated chemistry) and $\mathrm{CH_{4}}$ absorption
(carbon-dominated chemistry) because the relative strengths of the
1.15 and 1.4~$\mathrm{\mu m}$ absorption features are vastly different
for water and methane. Methane absorption bands at 1.15 and 1.4~$\mathrm{\mu m}$
are similarly strong and would inevitably result in two similarly
strong absorption features at 1.15 and 1.4~$\mathrm{\mu m}$, which
is not in agreement with the data.

Finding a lower limit on the C to O ratio is not possible because
no carbon-bearing species can be inferred from the data and a wide
range of water abundances is consistent with the detected water absorption
feature (Fig. \ref{fig:Water-abundance}). The detection of a single
absorption feature in low-resolution transmission spectroscopy generally
provides weak constraints on the abundance of that absorber \citep{benneke_atmospheric_2012}.
The shape and depths of the absorption features are predominately
determined by the scale height and cloud properties, and are only
indirectly dependent of the absorber abundance.

\subsection{Water Abundance\label{sub:Water-Abundance}}

The water abundances of all eight hot Jupiters are consistent with
the value expected for solar composition (Figures \ref{fig:Water-abundance}).
Although previous studies \citep{madhusudhan_h2o_2014} have reported
low water abundances and/or high carbon-to-oxygen ratios on HD~209458b,
HD~189733b, and WASP-12b, we find that these conclusions are not
supported by the data once the full range of clouds is considered
in the retrieval modeling. It is worth noting that the uncertainty
in the water abundance is predominately driven by the strong correlation
between water abundance and cloud top pressure (Figures \ref{fig:Water-abundance}).
The relatively small depths of the water absorption features at $1.4\,\mathrm{\mu m}$
as compared to cloud free scenarios clear can equally well be explained
by scenarios with low water abundance and low-altitude clouds (240~mbar)
or by solar water abundance scenarios with clouds between 1 and 100~mbar
(Figure \ref{fig:Water-vs-Clouds}). The correlation between cloud
top pressure and water abundance explain why lack of clouds in the
retrieval modeling would lead to the conclusion of low water abundance
as discussed in the following section.

\subsection{Importance of Considering A Wide Range of Clouds\label{sub:Importance-of-Considering}}

\begin{figure*}[t]
\noindent \begin{centering}
\includegraphics[width=1\textwidth]{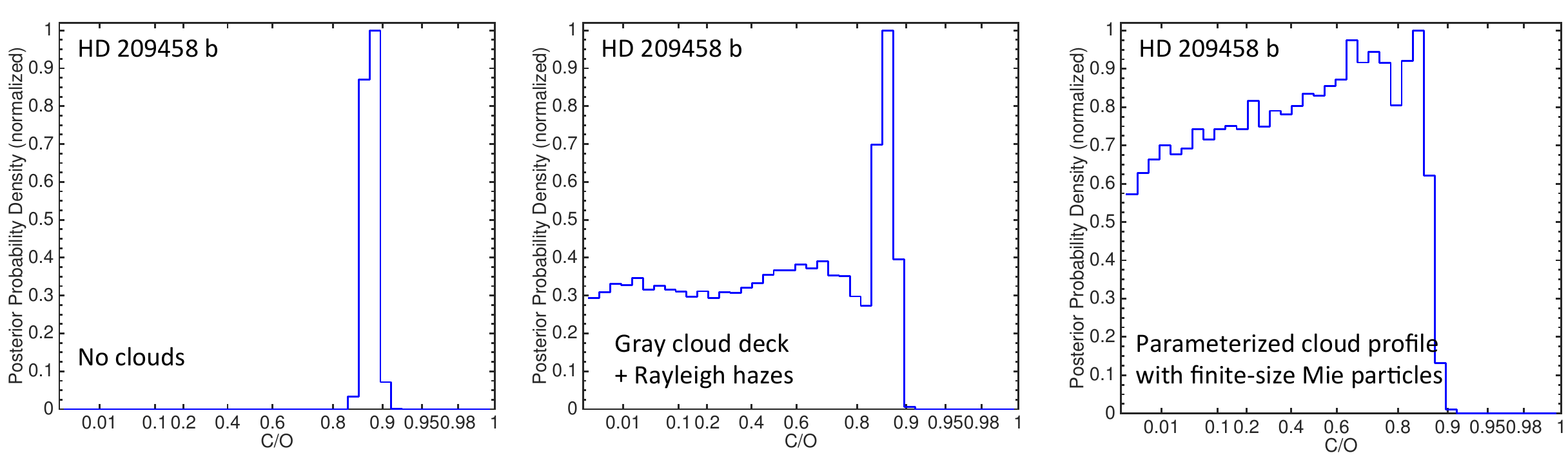}
\par\end{centering}

\noindent \centering{}\protect\caption{The constraints on the C/O ratio depend sensitively on the range of
clouds considered in the retrieval modeling. The three panels illustrate
the posterior probability density as function of C/O ratio for HD~209458b
assuming no clouds (left), accounting for a gray cloud deck and Rayleigh
hazes (center), and accounting for wide range of clouds using a parameterized
cloud profile and particle size distribution (right). The plotted
posterior probability distributions are marginalized over the model
parameters listed in Table \ref{tab:SCARLET-versus-regular-1-1}.
Completely ignoring the potential presence of clouds leads one to
the conclusion of high C/O ratio ($\mathrm{C/O=0.88\pm0.04}$) and
low water abundance (left). The peak for the retrieval model without
clouds (left) is narrow because only atmospheres with a narrow range
of water abundances corresponding to $\mathrm{C/O\approx0.88}$ fit
the relatively small water feature in the absence of clouds. However,
only an upper limit at C/O=0.89 can be inferred when clouds are considered
in the modeling. For the gray cloud deck + Rayleigh hazes model (center),
a peak at C/O=0.89 remain, but a long probability tail towards $\mathrm{C/O\rightarrow0}$
exists (center). No lower bound on C/O available. The peak at $\mathrm{C/O\approx0.88}$
also disappears in the parameterized cloud profile and particle size
model (right). The C/O results presented in this work use this parameterized
cloud profile and particle size model (right) to fully account for
the wide range of plausible clouds on hot Jupiters. \label{fig:DependencyOnClouds}}
\end{figure*}
The constraints on the atmospheric composition depend sensitively
on the range of clouds and hazes considered in the retrieval model.
It is therefore important to account for all plausible cloud scenarios
when retrieving the atmospheric composition. As an example, Figure
\ref{fig:DependencyOnClouds} compares the C to O ratio constraints
for HD~209458b for three different clouds models. Without the presence
of clouds, the only explanation for the relative shallow water absorption
feature in the\textit{ WFC3} observations of HD~209458b (Fig. \ref{fig:Model-transmission-spectra})
is a sub-solar water abundance and high C/O ratio (Figure \ref{fig:DependencyOnClouds}
left). The presence of clouds, however, presents an alternative explanation
for the observed shallow absorption feature, resulting in much weaker
constraints on C/O (Figure \ref{fig:DependencyOnClouds} center and
right). The sensitivity of the atmospheric composition constraints
is a result of strong degeneracies between the atmospheric composition
and cloud properties in low resolution transmission spectra (Figure
\ref{fig:Water-vs-Clouds}).

\subsection{Cloud and Haze Properties\label{sub:Clouds-and-Hazes}}

The observed transmission spectra indicate the presence of optically
thick cloud decks in the mid-atmospheres of HD~209458b and WASP-12b
(Section \ref{sub:Cloud-deck-on}). The clouds decks are found consistently
using both the ``cloud deck + Rayleigh haze''model as well as the
``parameterized cloud profile and particle size'' model. HD~189733b
may host a similar gray cloud deck if star spots affect the spectrum
(Section \ref{sub:HD189733b:-Small-Particle-Hazes}); otherwise thin
haze are required as previously suggested. No clouds or hazes are
inferred for WASP-19b, XO-1b, WASP-17b, and WASP-43b.

\subsubsection{Cloud Decks on HD~209458b, WASP-12~b, and WASP-31b\label{sub:Cloud-deck-on}}

\begin{figure*}[t]
\noindent \begin{centering}
\includegraphics[width=1\columnwidth]{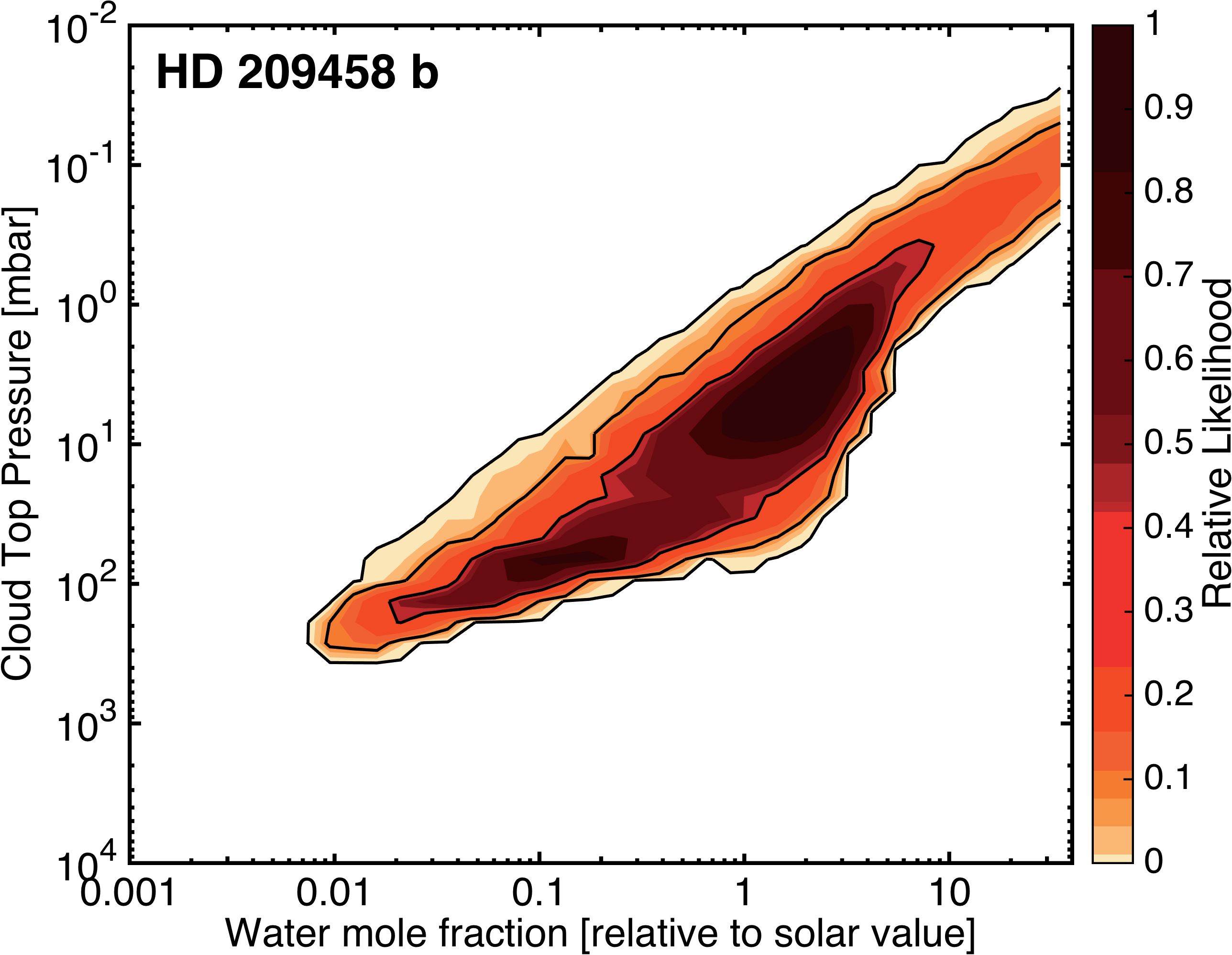}\includegraphics[width=1\columnwidth]{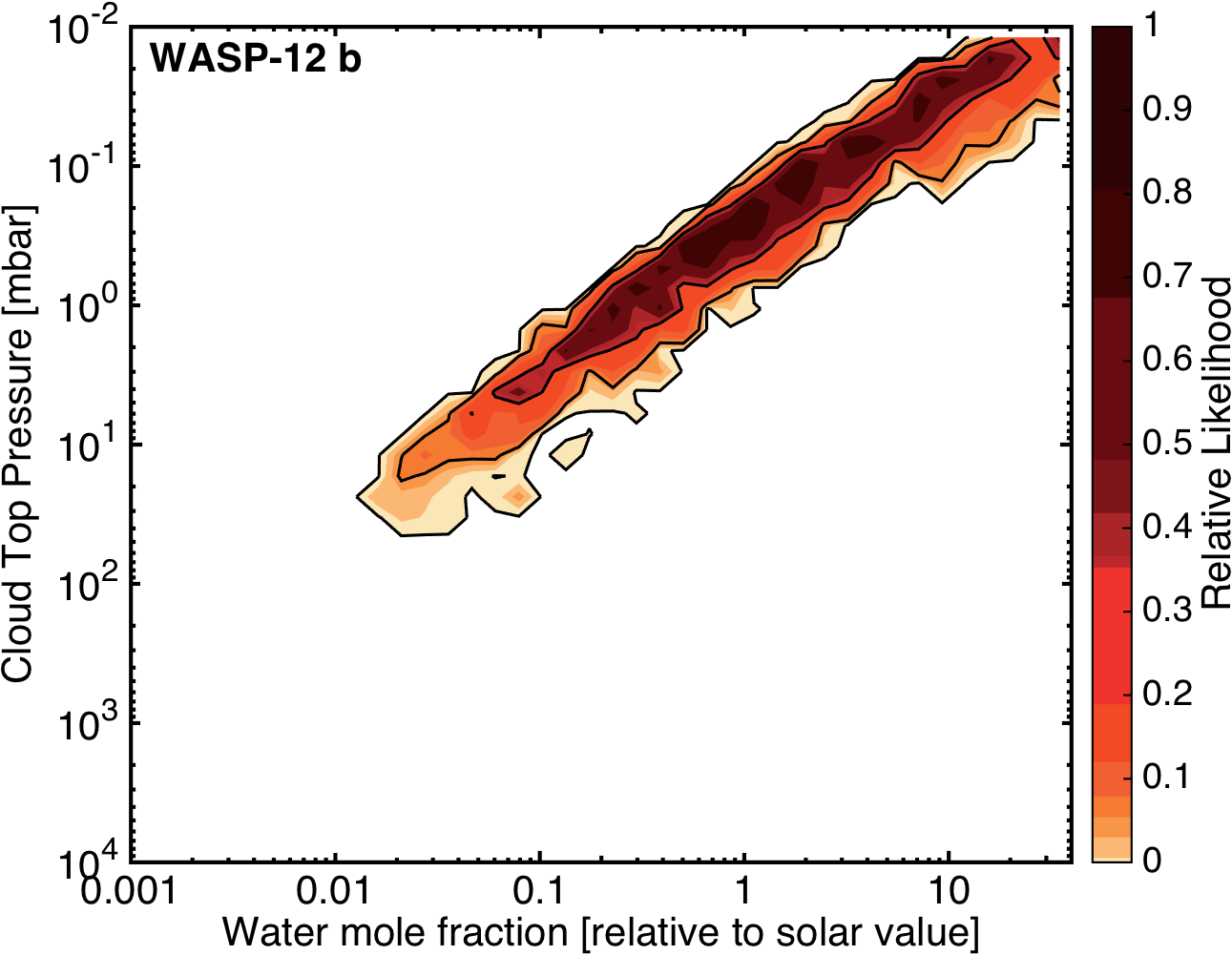}
\par\end{centering}

\noindent \begin{centering}
\includegraphics[width=1\columnwidth]{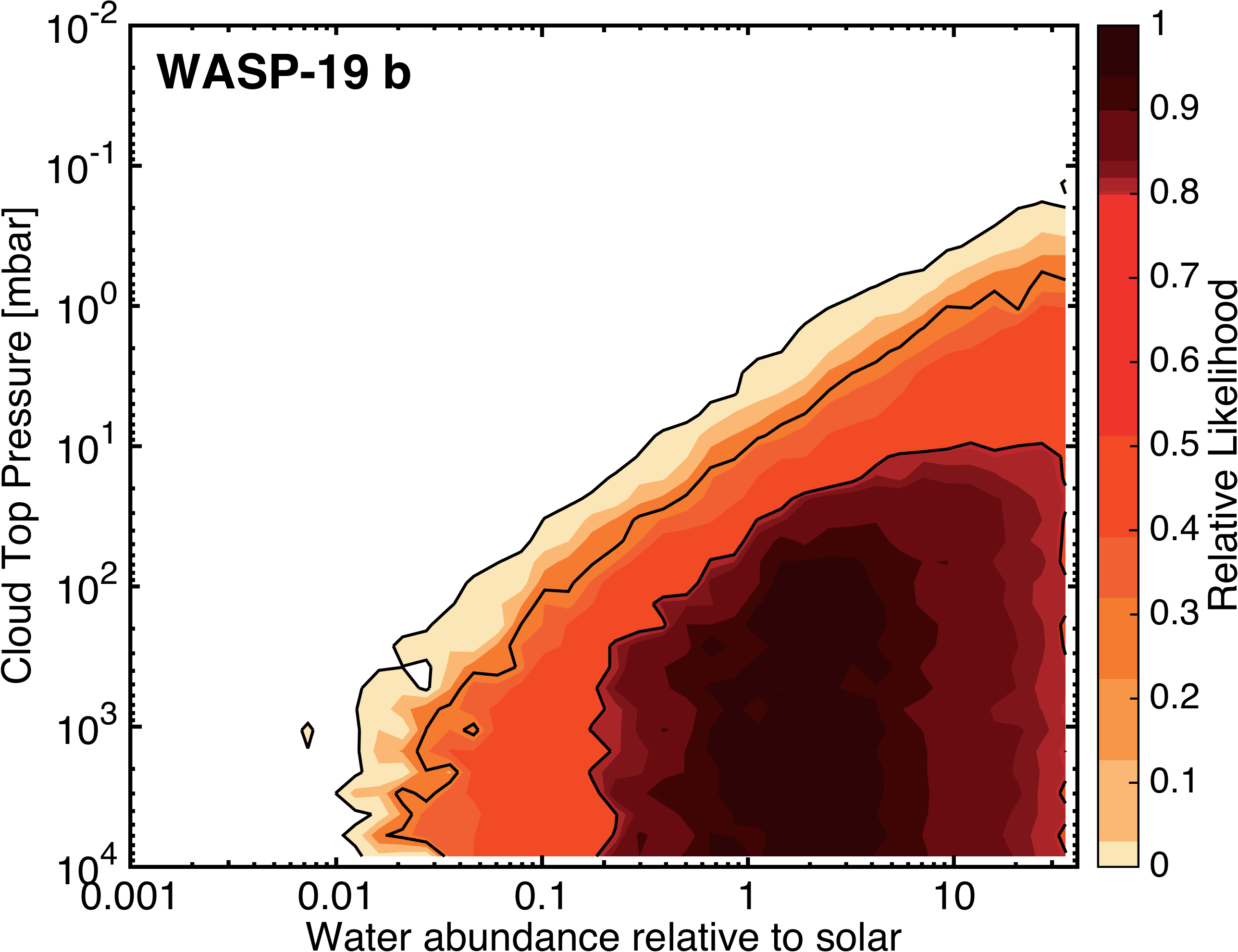}\includegraphics[width=1\columnwidth]{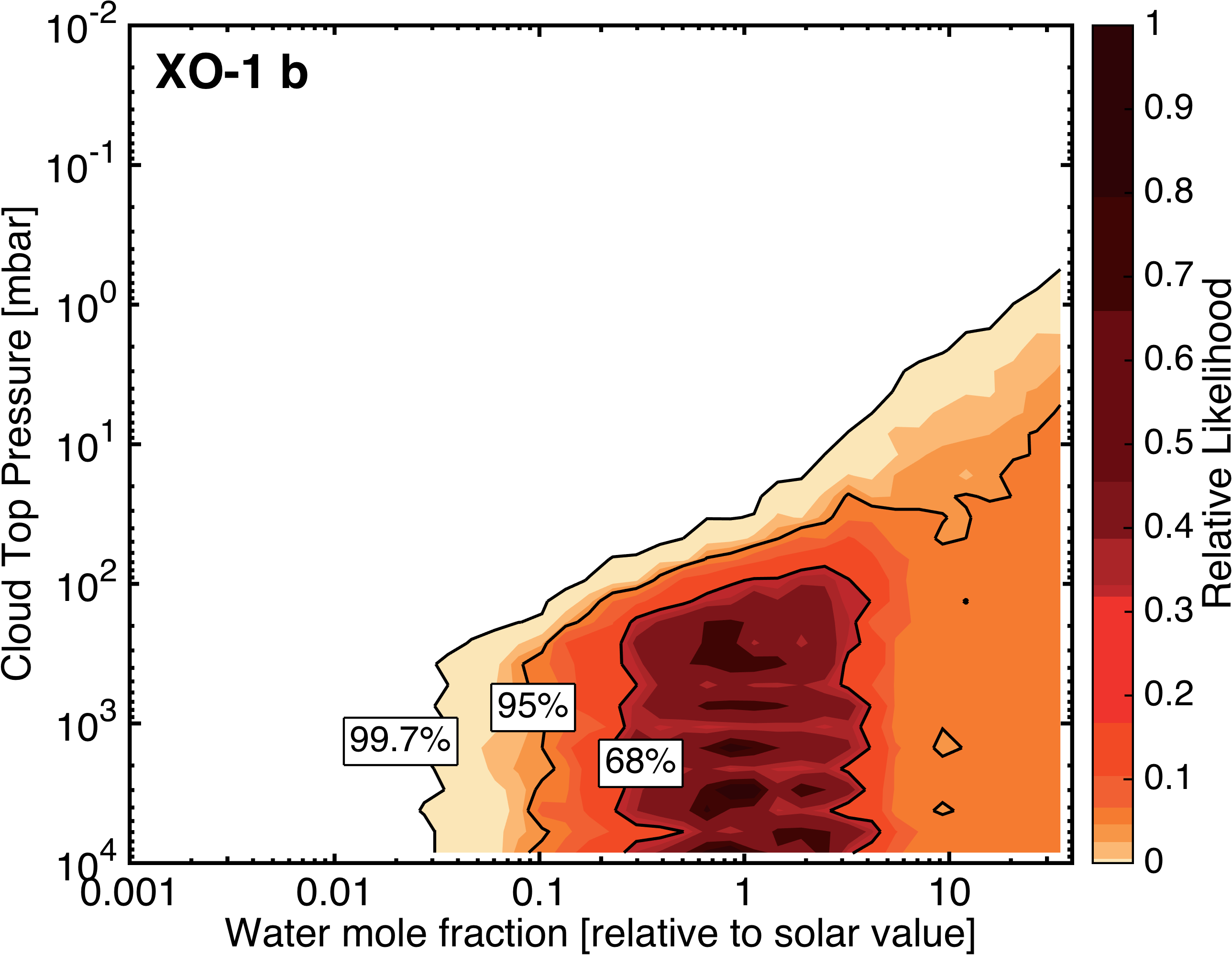}
\par\end{centering}

\noindent \centering{}\protect\caption{Joint constraints on the water abundance and cloud top pressure for
HD~209458b, WASP-12b, WASP-19b, and XO-1b. The colored shading indicates
the relative likelihood of atmospheric models as a function of water
mole fraction and cloud top pressure. Black contours mark the 68\%
(1$\sigma$), 95\% (2$\sigma$), and 99.7\% (3$\sigma$) confidence
regions, assuming a log-uniform prior on the water mole fraction.
Scenarios with largely cloud-free atmosphere are near the bottom of
each panel; scenarios with high altitude clouds are at the top. Clouds
are required on HD209458b and WASP-12b at greater than 99.7\% confidence.
Water abundance and cloud top pressure are highly correlated, preventing
precise constraints on water abundance and cloud top pressure individually.
For WASP-19b and XO-1b, current observations do not distinguish between
cloud free scenarios and scenarios with low-altitude clouds. Scenarios
with and without clouds are in agreement with the data. The lower
limit on the cloud top pressure depends on the water abundance. \label{fig:Water-vs-Clouds}}
\end{figure*}
\begin{figure}[!t]
\noindent \begin{centering}
\includegraphics[width=0.8\columnwidth]{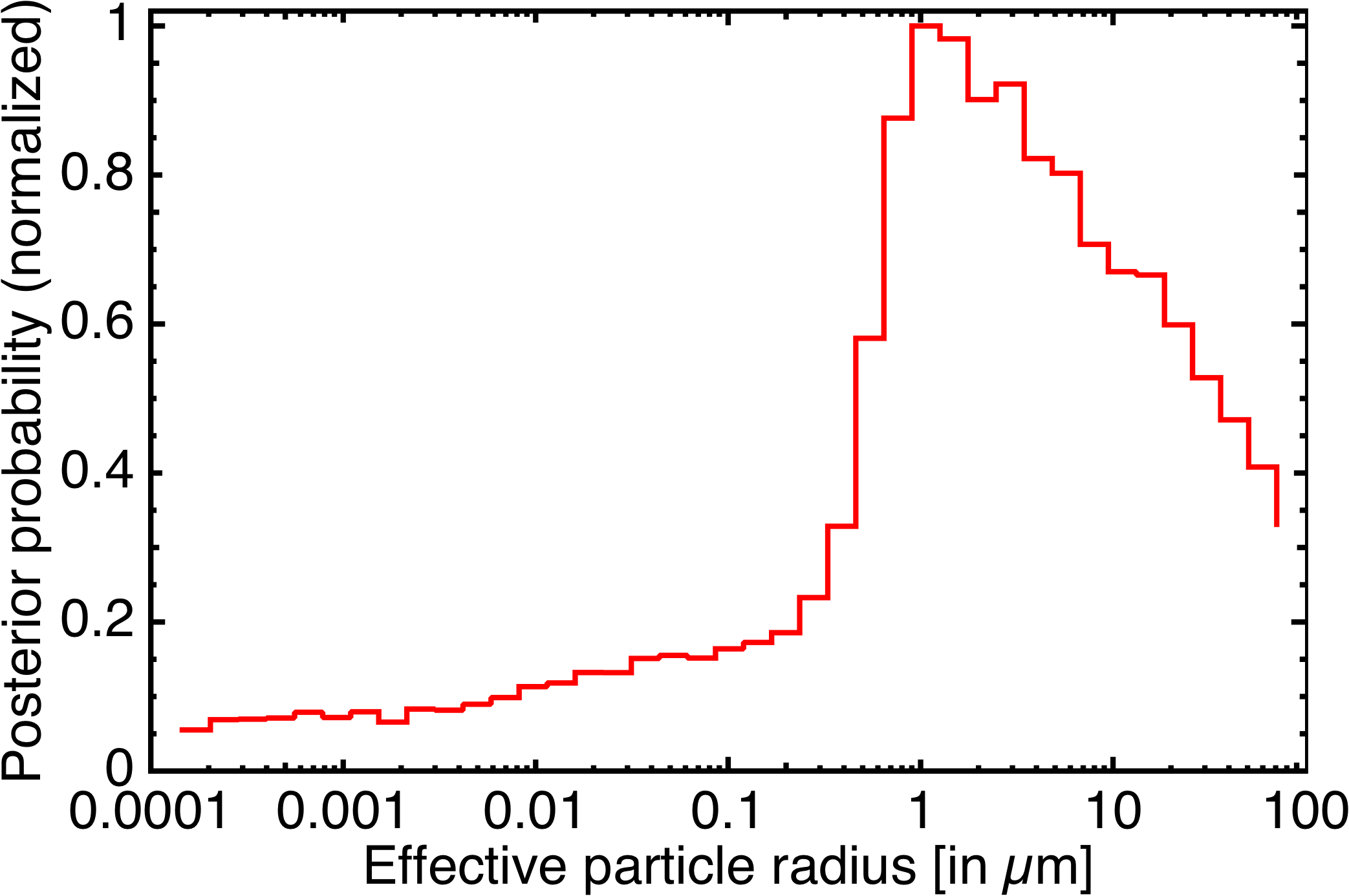}
\par\end{centering}

\noindent \begin{centering}
\protect\caption{Observational constraints on effective particle size of the clouds
on HD~209458b. The most probably scenarios for the clouds on HD~209458b
are large particles (>1~$\mu\mathrm{m}$). Smaller particles cannot
fully be ruled out, however, because a sharp upper cloud deck would
mask the strongly wavelength-dependent scattering properties of small
particles in transmission spectra (see also Figure \ref{fig:Particle-size-HD209458b}).
The assumed particles are Mie scattering particles with the complex
refractive indices of $\mathrm{MgSiO_{3}}$.\label{fig:Spectra-cloud-compositions}}
\includegraphics[width=0.4\textwidth]{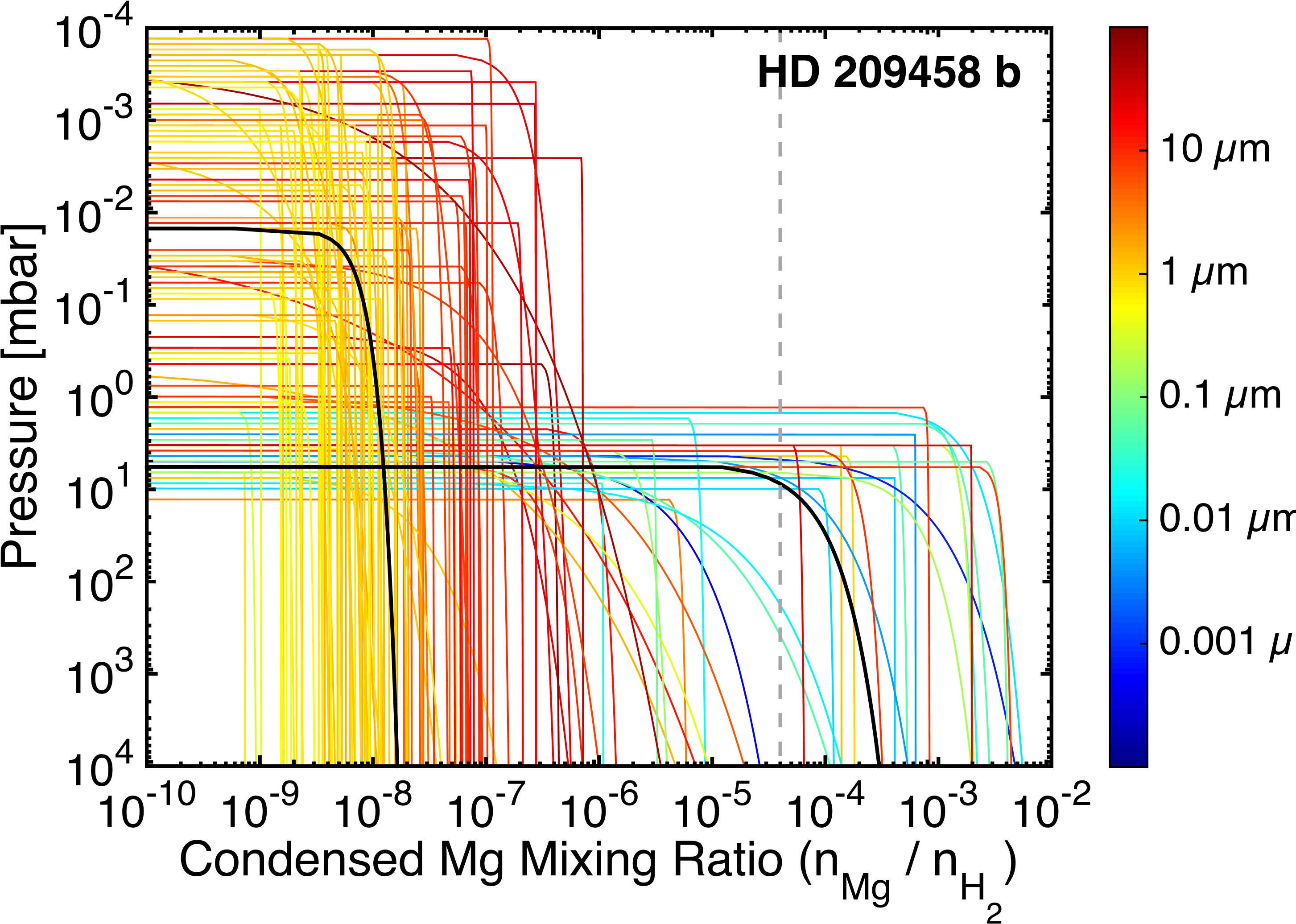}
\par\end{centering}

\noindent \centering{}\protect\caption{Observational constraints on the vertical cloud density profiles for
$\mathrm{MgSiO}_{3}$ grains in the atmospheres of HD~209458b and
HAT-P-1b. Representative cloud density profiles are plotted for scenarios
that are within a likelihood ratio of 300:1 ($\Delta\chi^{2}<11.8$)
of the best fit to the observations. For illustrative purposes, cloud
profiles are shown for near solar composition atmospheres only (0.5<M<5;
0.3<C/O<0.6). The cloud density is represented by the ratio between
the number of condensed $\mathrm{MgSiO}_{3}$ molecules and the number
of background $\mathrm{H_{2}}$ molecules. The solar abundance number
ratio $\mathrm{Mg/H=}4\cdot10^{-5}$ is indicated as a vertical dashed
line. Colors indicate the effective particle radii for each of the
cloud profiles. Two distinct types of clouds scenarios are in agreement
with the observations of HD~209458b: 1) a sharp cloud deck near 1-10~mbar
with a cloud free upper atmospheres or 2) thin clouds composed of
large particles (>1~$\mu\mathrm{m}$) reaching to high altitudes.
Both scenarios reduce the amplitude of NIR molecular feature in a
largely wavelength-independent way. In contrast, HAT-P-1b must host
thin, small-particle ($\lesssim0.1\,\mu\mathrm{m}$) reaching to high
altitude (>1~mbar) to explain the enhanced opacity at visible wavelength.
\label{fig:Particle-size-HD209458b}}
\end{figure}

\begin{figure}[t]
\includegraphics[width=1\columnwidth]{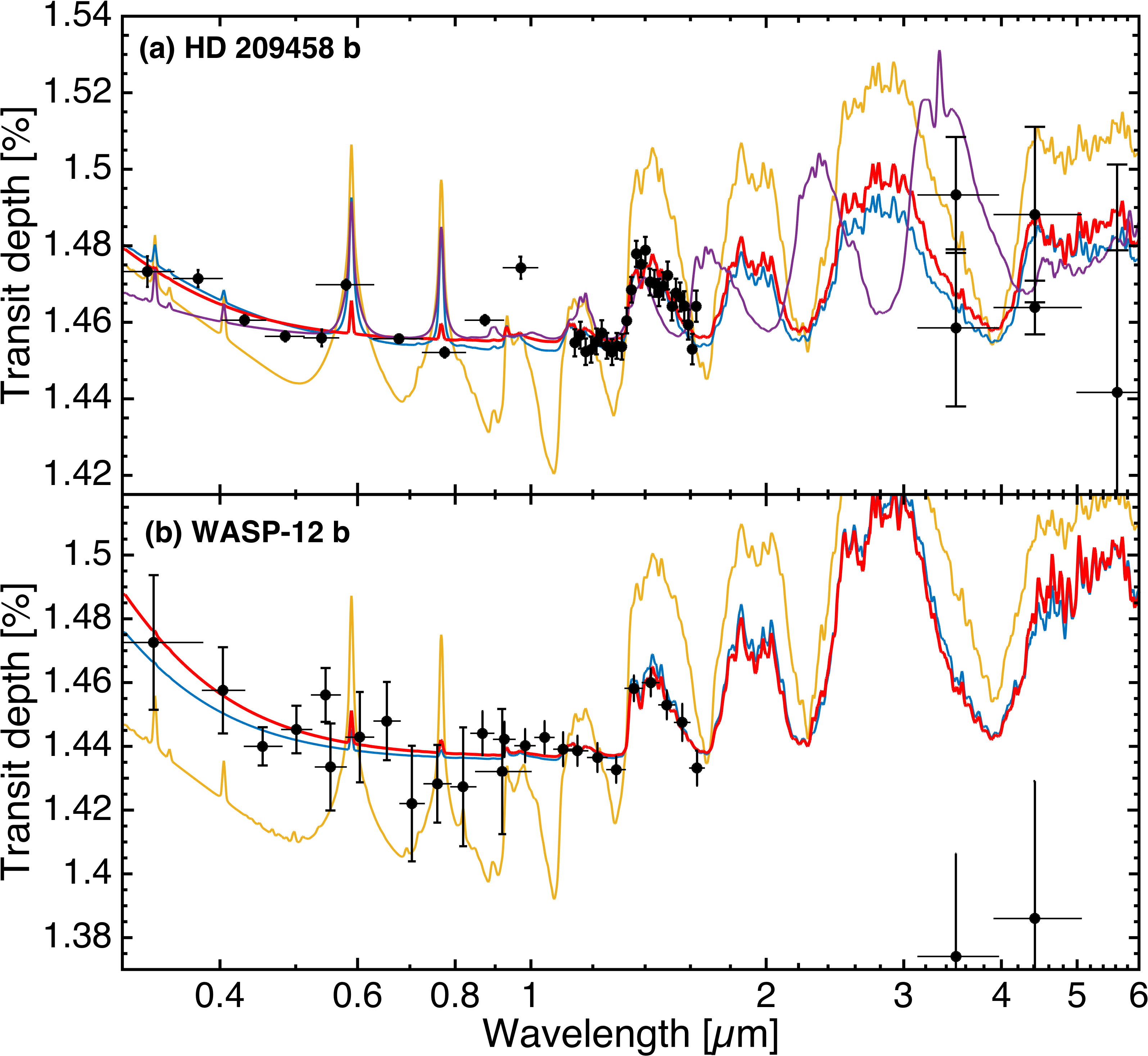}

\protect\caption{Ultraviolet to near-infrared transit depth measurements of HD~209458b
and WASP-12b compared to model transmission spectra. Solid lines show
the transmission spectra for a fiducial clear solar composition atmosphere
(yellow), the overall best-fitting model (red), the best-fitting solar
composition atmosphere with clouds (blue). Black circles indicate
the transit depth measurement and their 1$-\sigma$ uncertainties
taken from \citet{knutson_using_2007},\citet{deming_infrared_2013},
\citet{evans_uniform_2015}, and \citet{crossfield_spitzer/mips_2012}
for HD~209458b, and\citet{sing_hst_2013} and\citet{kreidberg_detection_2015}
for WASP-12b. The observed transmission spectra indicate the presence
of optically thick cloud decks on all three planets in the mid-atmospheres
of HD~209458b and WASP-12b. Only the strong $\mathrm{1.4\,\mu m}$
water band, the cores of alkali lines, and the Rayleigh scattering
signature blue of 500~nm are detectable.\label{fig:UV-to-IR}}
\end{figure}

It is instructive to investigate the constraints on cloud top pressure
and haze opacity from the simple ``gray cloud deck + Rayleigh hazes''
model. This two-parameter cloud model simultaneously allows for a
gray cloud deck at a parameterized cloud top pressure as well as ``Rayleigh''
hazes with parameterized opacity composed of small particles ($r_{p}\ll\lambda$).
Within this model, the \textit{HST} \textit{WFC3} observations of
HD~209458b and WASP-12b strongly indicate the presence of a thick
cloud deck. All scenarios without clouds are ruled out at greater
99.7\% confidence (Figures \ref{fig:Water-vs-Clouds}). Based on the
depth of the observed water feature size, I infer that the upper cloud
deck is in the mid-atmosphere between approximately 200~mbar and
0.01~mbar for HD~209458b and 30~mbar and 0.01~mbar for WASP-12b.
Stronger constraints on the cloud top pressure are not available because
the cloud top pressure is strongly correlated with the water mole
fraction (Figure \ref{fig:Water-vs-Clouds}). For HD~209458b, equally
good fits to the data are obtained for a wide range of atmospheric
scenarios ranging from low water abundances (1\% solar) and low-altitude
clouds (200~mbar) to high water abundances (10~x~solar) and high-altitude
clouds (0.01~mbar). The clouds in the atmosphere of WASP-12b are
between 0.01~mbar for 5\% solar water abundance and 10~mbar for
10 times the solar water abundance (Figure \ref{fig:Water-vs-Clouds}).
The inferred clouds top pressure for a given water mole fraction is
lower for WASP-12~b because the observed water absorption feature
is weaker as compared to the cloud free model transmission spectrum
than for HD~209458b (Figure \ref{fig:Model-transmission-spectra}).
The presence of clouds can inferred from the WFC3 observations alone,
but is also strongly supported by the HST STIS observations (Figure
\ref{fig:UV-to-IR}).

The negative correlations between water mole fraction and cloud top
pressures in Figure \ref{fig:Water-vs-Clouds} can be explained as
follows. High clouds reduce the depths of near-IR water absorption
features. In contrast, increasing the water abundance strengthens
the water absorption features for a given cloud top pressure. Different
combinations of water mole fraction and cloud top pressure, therefore,
result in the same depths of the 1.4~$\mu\mathrm{m}$ water absorption
feature and similar good fits to the data. 

Modeling the cloud particles using Mie scattering reveals that particles
with effective diameters above $1\,\mu\mathrm{m}$ are most probable
(Figure \ref{fig:Particle-size-HD209458b}). Large particles naturally
explain the gray nature of the observed cloud signature. Both the
1.15 and 1.4~$\mu\mathrm{m}$ water absorption features are diminished,
as expected for a sharp, wavelength-independent cutoff of the grazing
star light below the cloud top pressure. It is interesting to note,
however, that small particles can similarly appear gray in exoplanet
transmission spectra despite their non-gray scattering properties,
if the particle density increases sharply near the cloud top. A sudden
rise in particle number density at a given altitude can lead to the
cutoff of all grazing light beams below that altitude, largely independent
of the wavelength. The effect of cloud particles on exoplanet transmission
spectra depends not only on the scattering properties of the particles,
but also sensitively on the vertical distribution of the particles
in the atmosphere.

One caveat for HD~209458~b is that the $\mathrm{1\,\mu m}$ data
point at the red end of the HD209458b \textit{HST STIS} observations
cannot be fit by any model spectrum. Water, TiO, and VO are plausible
absorbers at 1~$\mu\mathrm{m}$, but cannot account for the increased
transit depth without also increasing the transit depth in the surrounding
\textit{HST STIS} and \textit{WFC3} transit depth measurements. The
increased transit depth measurement may either be caused by an unknown
opacity source or systematic effects potentially related to the fringing
of the STIS CCD may have affected the measurement.

Similarly, the Spitzer measurement at 3.6 and $\mathrm{4.5\,\mu m}$
cannot be fit to better than $1.5-\sigma$ when simultaneously fitting
the \textit{HST STIS}, \textit{WFC3}, and \textit{Spitzer} data points
(Figure \ref{fig:UV-to-IR}b). While the $1.5-\sigma$ deviation of
two points does not substantially degrade the overall statistical
fit to the data; it is worth a discussion because they are the only
measurements red of 3~$\mathrm{\mu m}$. \citet{sing_hst_2013} and
\citet{stevenson_deciphering_2014} suggested that the entire UV to
near-IR spectrum is sloped due to small particle hazes. However, the
new \textit{HST WFC3} data by \citet{kreidberg_detection_2015} rule
out this scenario unless they are substantially offset to the\textit{
HST STIS} data. No plausible model exists that would simultaneously
fit the reported \textit{HST STIS}, \textit{WFC3}, and \textit{Spitzer}
data points to within better than $1.5-\sigma$. If water absorption
increases the transit depth to 1.46\% at $1.4\,\mu\mathrm{m}$, then
the absorption of the same water vapor would result in a transit depth
between 1.44\% and 1.5\% in the Spitzer bandpasses, independent of
the cloud or haze properties. Possible explanation are that the transit
depth measurements are low due to random chance or that they are affected
by star spots \citep{mccullough_water_2014}, WASP-12~b's stellar
companion \citep{stevenson_deciphering_2014}, stellar variability,
or uncorrected instrumental effects.
\begin{figure}[t]
\includegraphics[width=1\columnwidth]{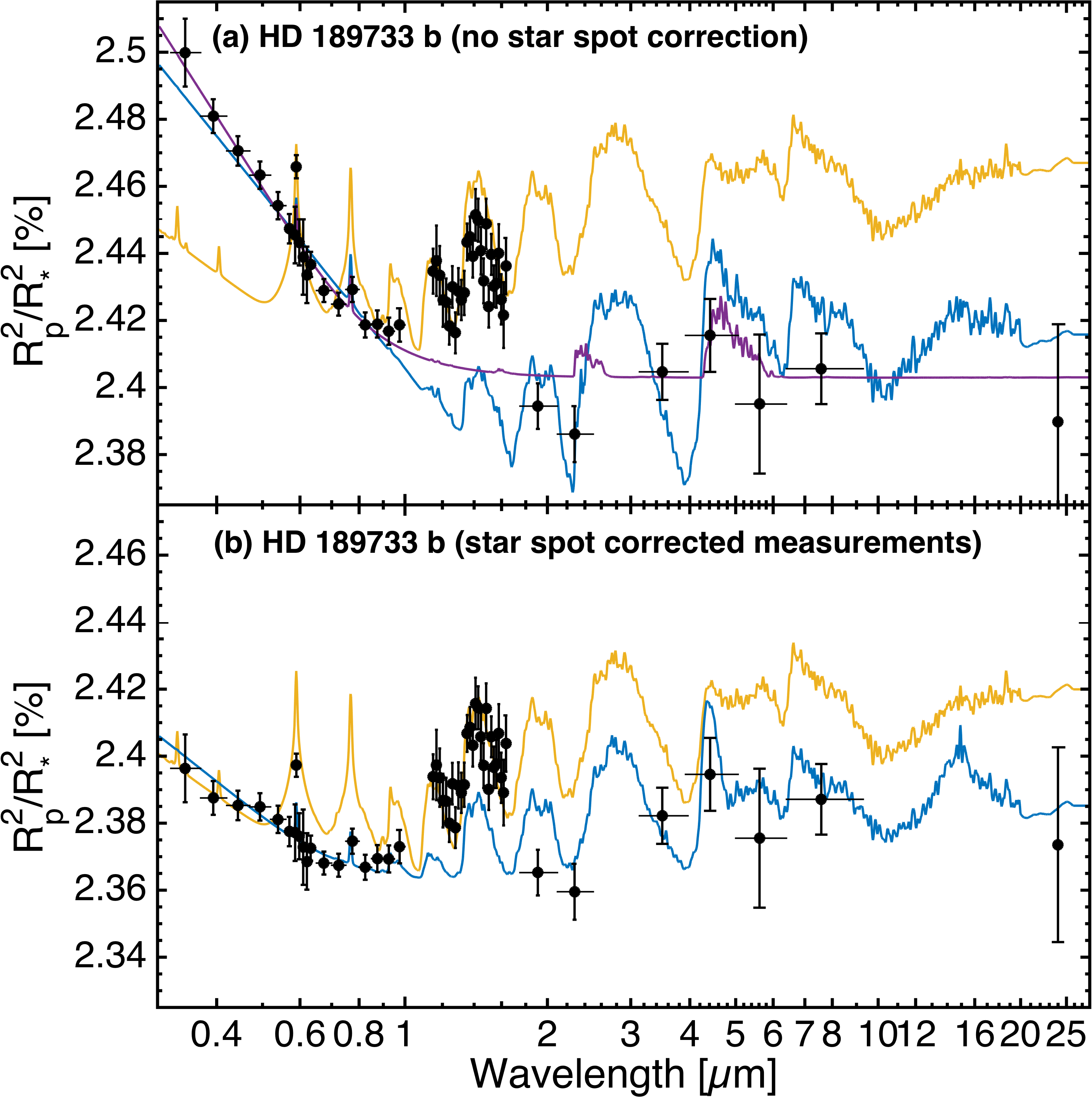}

\protect\caption{Ultraviolet to near-infrared transit depth measurements of HD~189733b
compared to model transmission spectra. Black circles in panel (a)
indicate the transit depth measurements directly as reported by \citet{pont_prevalence_2013}
and \citet{mccullough_water_2014}. Panel (b) shows $\left(R_{P}/R_{*}\right)^{2}$
based on the same measurements, but corrected for the presence of
star spots according to \citet{mccullough_water_2014} and assuming
the best fitting star spot fraction and spot temperature. The star
spot corrected measurements reveal that the spectroscopic appearance
of HD~189733b can be similar to the one of HD~209458b and WASP-12b
(Figure \ref{fig:UV-to-IR}). Solid lines show the transmission spectra
for a fiducial clear solar composition atmosphere (yellow), the best-fitting
solar composition atmosphere with clouds and hazes (blue), and a theoretical
model with only a cloud deck and high altitude hazes (purple). The
originally reported transit depth measurements (panel a) strongly
indicate the presence of high altitude, small-particle hazes to match
the Rayleigh slope between 0.3 and 1. However, an significant offset
of the \textit{WFC3} data is required to match the data. After star
spot correction (panel b), the measurement do not favor hazy atmospheric
scenario, but are in better agreement with either a completely cloud
free atmosphere with Na and K removed or scenarios with a thick cloud
deck at 1-100~mbar similar to HD~209458b. Cloud-free scenarios without
Na and K simultaneously match the \textit{HST STIS }and WFC3 with
all Spitzer significantly offset. Cloudy scenarios matches the \textit{Spitzer}
and \textit{HST STIS} observations with the \textit{WFC3} offset.
Plausible explanations for the suggested offsets are variability in
unocculted star spot which would give additional support to the star
spot hypothesis as an explanation for the short wavelength slope.
\label{fig:UV-to-IR_HD189}}
\end{figure}

WASP-17b and WASP-43b show water absorption features weaker than expected
for a cloud free solar composition atmosphere (Figure \ref{fig:Model-transmission-spectra}),
which may indicate the presence of clouds similar to HD~209458b.
However, the data is insufficient to rule out low water abundance
scenarios without clouds for WASP-17b and WASP-43b. 

The observations of WASP-19b and XO-1b do currently not indicate the
presence of clouds or hazes. Instead, the observations provide a strict
lower limit on the cloud top pressure and a strict upper limit on
the opacity of small-particle ``Rayleigh'' hazes in the upper atmosphere.
For WASP-19b, no cloud deck exists above the 10~mbar level for solar
water abundance and 0.1~mbar for 50 x solar (Figure \ref{fig:Water-vs-Clouds}).
Similar, if small particle hazes are present, their opacity does not
exceed $\sigma_{R,400}=10^{-4}\,\mathrm{m^{2}/kg}$ for a solar metallicity
atmospheres and $,\sigma_{R,400}=1\,\mathrm{m^{2}/kg}$ for a 50~x~solar
metallicity atmospheres. Higher values for haze opacity would mute
the $\mathrm{H_{2}O}$ absorption features in the \textit{HST WFC3}
bandpass and result in a sloped visible and NIR transmission spectrum,
which is not in agreement with the observations. Qualitatively the
results are similar for XO-1b (Figure \ref{fig:Water-vs-Clouds}).

\subsubsection{HD~189733b: Small-Particle Hazes or Gray Cloud Deck?\label{sub:HD189733b:-Small-Particle-Hazes}}

The reported transmission spectrum of HD~189733b shows a prominent
slope at short-wavelength with increasing transit depth from the near-IR
(1~$\mu\mathrm{m}$) to the ultraviolet (300~nm) (Figure \ref{fig:UV-to-IR}a).
This slope has traditionally been attributed to scattering of non-absorbing,
Rayleigh scattering dust \citep{lecavelier_des_etangs_rayleigh_2008,sing_hubble_2011,pont_prevalence_2013}.
Recently, \citet{mccullough_water_2014} suggested that unocculted
star spots may be responsible for the apparent slope in the transit
depth, reinterpreting the spectrum of HD~189733b's with a cloud free
atmosphere. In this section, I revisit the observed spectrum by (1)
assuming that the reported measurements of the planet-to-star radius
ratios are not affected by star spots, and (2) accounting for the
presence of unocculted star spots. In the presence of star spots,
I find that HD~189733b may have a cloud deck in the mid-atmosphere
similar to HD~209458b, albeit at higher pressure. If confirmed, the
cloud properties of HD~189733b and HD~209458b could be more similar
than previously believed.

First, assuming that the reported measurements are not affected by
star spots, I find that small-particle ``Rayleigh'' hazes significantly
exceeding the opacity of molecular Rayleigh scattering must be present
in the upper atmosphere of HD~189733b, in agreement with previous
studies. Assuming a Rayleigh-like haze opacity ($\sigma\propto\lambda^{-4}$),
the reference opacity at 400~nm, $\sigma_{R,400}$, is between $\sim10^{-2}\,\mathrm{m^{2}kg^{-1}}$
for low metallicity atmospheres and $\sim10\,\mathrm{m^{2}kg^{-1}}$
for 50~x~solar composition atmospheres. The inferred haze opacity
exceeds the expected opacity by molecular Rayleigh scattering by a
factor of 10 to 10,000. A strong correlation between haze opacity
and atmospheric metallicity arises because different combinations
of haze opacity and atmospheric metallicity result in identical relative
transit depths between the visible and infrared spectrum. Increased
metallicity increases transit depth in the near-IR, but the same relative
transit depth between near-IR and visible wavelengths can be recovered
if the haze opacity is increased simultaneously \citep{benneke_atmospheric_2012}.

Alternatively, assuming that star spots affect the measured transmission,
I correct the apparent transit depth measurement using the relation
\begin{equation}
\left(\frac{R_{p}}{R_{*}}\right)^{2}=D\cdot\left[1-\delta\left(1-\frac{e^{hf/kT_{phot}}-1}{e^{hf/kT_{spot}}-1}\right)\right],
\end{equation}

where $D$ is the measured transit depth, $R_{p}/R_{*}$ is the geometric
planet-to-star radius ratio, $\delta$ is the star spot fraction,
$T_{\mathrm{spot}}$ is the star spot temperature, $T_{phot}$ is
the effective temperature of the star's photosphere, and $f=c/\lambda$
is the observing frequency \citep{mccullough_water_2014}. The corrected
transit depth, $(R_{p}/R_{*})^{2}$, is illustrated in Figure \ref{fig:UV-to-IR_HD189}(b)
assuming the best fitting spot fraction $\delta=4.6\%$ and spot temperature
$T_{spot}=4400\, K$. After spot correction, the transmission spectrum
of HD~189733b shows similar trends to HD~209458b and WASP-12b. Similar
to HD~209458b and WASP-12b, a gray cloud deck may be present on HD~189733b
that reduces the reduced the depth of the water absorption feature
at $\mathrm{1.4\,\mu m}$ and mute the wide wings of potassium and
sodium. Molecular Rayleigh scattering in the high atmosphere is visible
at wavelength shorter than 500~nm.

\section{SUMMARY \& CONCLUSIONS}

I have introduced a novel approach to interpret spectroscopic observations
of planetary atmospheres that combines the statistical robustness
and exploratory nature of atmospheric retrieval methods \citep{rodgers_inverse_2000,madhusudhan_temperature_2009,benneke_atmospheric_2012,line_systematic_2013}
with the self-consistency and ability to provide physical insights
of complex atmospheric forward models \citep{burrows_nongray_1997,burrows_theoretical_2005,marley_clouds_2002,fortney_synthetic_2008,fortney_self-consistent_2011,moses_disequilibrium_2011-1}.
Rather than retrieving only molecular abundances as free parameters,
the new ``SCARLET'' approach can provide direct insights into the
elemental composition of the deep atmosphere based on observations
of the upper atmosphere. 

In this work, I applied the new SCARLET framework to eight transiting
hot Jupiters with detected molecular absorption features in their
near-infrared\textit{ }transmission spectra (HD~209458b, WASP-19b,
HAT-P-1b, XO-1b, HD~189733b, WASP-12b, WASP-17b, and WASP-43b). My
main findings are as follows:
\begin{enumerate}
\item The C/O ratios in the deep atmospheres of all eight hot Jupiters are
consistently below 0.9. This result strongly indicates that gas envelopes
of most hot Jupiter are not carbon-dominated (C/O>1). The finding
of $\mathrm{C/O\lesssim0.9}$ is robust for HD~209458b, WASP-19b,
WASP-12b, HAT-P-1b and XO-1b (Table \ref{tab:SCARLET-versus-regular-1-1}).
A comprehensive exploration of the parameter space reveals that any
scenario with $\mathrm{C/O>1}$ is excluded at high significance.
All carbon-rich scenarios (C/O>1) for HD~209458b, for example, are
excluded at $\Delta\chi^{2}>55.9$ (likelihood ratio of $1.2\cdot10^{12}$
to 1). The finding of $\mathrm{C/O\lesssim0.9}$ relies solely on
\textit{HST WFC3 }transmission spectroscopy, which has been shown
to provide repeatable and reproducible measurements \citep[e.g.,][]{kreidberg_clouds_2014}.
No lower limit is provided because all C/O ratios smaller than 0.9
are equally plausible for all eight planets.
\item The water abundances for all eight planets are consistent with the
value expected for a solar composition. For HD~209458b, I find that
the relatively small depth of the 1.4~$\mathrm{\mu m}$ water absorption
feature as compared to a clear solar composition atmosphere is due
to the presence of a cloud deck, not due to a low water abundance
as suggested by \citet{madhusudhan_h2o_2014}. In general, the constraints
on the water abundance remain weak and are driven by the strong correlations
between the water abundance and cloud top pressure.
\item Current observations of transiting hot Jupiters provide no meaningful
constraints on the overall abundance of metals in the atmospheres.
All metallicity values in the probed range between 0.1 and 100 times
solar metallicity are in good agreement with the observed transmission
spectra. \citep{benneke_atmospheric_2012}. HD~189733b may host a
similar gray cloud deck if star spots affect the spectrum. 
\item Clouds on hot Jupiters are common over a wide range of equilibrium
temperatures. The transmission spectra of HD~209458b $(T_{\mathrm{eq}}=1400\,\mathrm{K})$
and WASP-12~b $(T_{\mathrm{eq}}=2500\,\mathrm{K})$ strongly indicate
the presence of cloud decks in the mid-atmosphere. The cloud top pressure
is between 0.01~mbar and 200~mbar at 99.7\% confidence, strongly
correlated with the atmospheric water abundance. HD~189733b may host
a cloud deck similar to HD~209458b and WASP-12~b, rather than the
suggested Rayleigh hazes, if star spots affect the observed spectrum.

\item An important lesson from this study is that the constraints on the
atmospheric composition depend sensitively on the range of clouds
and hazes considered in the retrieval model. Ignoring the presence
of clouds or describing the clouds with a more restricted model may
produce overly optimistic constraints on the atmospheric compositions.
In this work, I account for a broad variety of cloud properties and
derive reliable constraints on composition by marginalizing over free
parameters describing the particle size and density profile of the
clouds.
\item The posterior probability distributions of atmospheric parameters
obtained from low signal-to-noise exoplanet spectra are highly non-Gaussian.
Parameters, such as the C/O ratio, may have only one-sided constraints
or long tails at low probability. In this regime, presenting the traditional
$\pm1-\sigma$ (68\%) uncertainties can be extremely misleading because
long tails reaching to values far from the reported best-fit may substantially
change the interpretation of the results. In this work, I discuss
99.7\% confidence limits whenever possible and provide likelihood
ratios to assess our confidence in excluding atmospheric scenarios.
\end{enumerate}

\subsection{Implications for Planet Formation}

The upper limit of 0.9 on the C/O ratio has important implications
for the formation of hot Jupiters. Protoplanetary disk models suggest
that oxygen near the disk's mid-plane freezes out efficiently beyond
the water ice-line, resulting in oxygen-depleted gas between water
ice line and CO lines \citep{oberg_disk_2011}. \citet{helling_disk_2014}
showed that the protoplanetary gas between water and CO ice lines
should transition to $\mathrm{C/O\rightarrow1}$ on timescales of
$\sim3\,\mathrm{Myrs}$ set by cosmic-ray induced unblocking of $\mathrm{O_{2}}$
and CO. Assuming that most of the giant planet envelope directly accretes
from this protoplanetary gas, the formation of giant planets with
carbon-enriched envelopes would be a natural consequence. 

The observed upper limit presented here implies that either the gas
disk has not fully transitioned to $\mathrm{C/O\rightarrow1}$ at
the time that the planet undergoes runaway accretion or the planet's
gas envelope is subsequently strongly polluted by the late accretion
of leftover planetesimals. Alternatively, most of the metals in the
gaseous envelope may be delivered before the planet undergoes runway
accretion. In van Boekel et al., (2015, in preparation), we find that
the core of a giant planet can acquire a gaseous envelope well before
the onset of runaway gas accretion. Once this envelope is sufficiently
thick for infalling planetesimals to evaporate before reaching the
planet's core, the planet's core stops growing and any accreted planetesimals
contribute to the growth of a metal-rich gas envelope. Runaway accretion
of metal-poor protoplanetary gas will eventually dilute this initial
envelope with vast amount of H/He; however, in this scenario the metal
composition would be more representative of the solids in the protoplanetary
disk than of the gas. A metal composition representative of the solids
in the protoplanetary disk would naturally explain a low C/O ratio.

Finally, the observed upper limit on the C/O ratio rules out any scenario
in which the planet forms inside the water snowline and accretes refractory
and volatile materials with ISM-like carbon abundances (van Boekel
et al., 2015, in preparation). Such scenarios would lead to C/O>1,
which is not in agreement with the findings presented here. Formation
within the water ice line remains possible, though, if the refractory
and volatile materials are oxygen-rich similar to refractory and ices
materials in the Solar System.

\subsection{Future Work}

Observations with wider spectral coverage or substantially more precision
and spectral resolution are needed to place stringent constraints
on the C/O ratios in hot Jupiter gas envelopes. Despite the limitations
of the current data, I was able to place a stringent upper limit of
0.9 on the C/O ratios of studied atmospheres due to the presence of
detectable water absorption in their transmission spectra. In the
future, the best way of also providing lower limits on the C/O ratios
is to also detect the absorption of carbon-bearing species such as
CO and $\mathrm{CH_{4}}$. Transmission spectra, in particular, provide
a straightforward way of determining the relative abundances of $\mathrm{H_{2}O}$,
$\mathrm{CO}$, and $\mathrm{CH_{4}}$ by comparing the relative transit
depths in the molecular absorption bands of these molecules \citep{benneke_atmospheric_2012}.
In this case, the SCARLET model can directly deliver the C/O ratio
from observations covering near-infrared $\mathrm{H_{2}O}$ and $\mathrm{CO}$
or $\mathrm{CH_{4}}$ absorption bands.

On the planet formation theory side, it will be essential to understand
whether the elemental compositions of giant planet envelopes are predominately
set by gas phase accretion \citep{oberg_disk_2011,helling_disk_2014}
or by accretion of solid and ices in the form of planetesimals (van
Boekel et al., 2015 in preparation). For that, it will be crucial
to understand the relative time scales for accretion of an initial
gas envelope and planetesimals as well as the initialization of the
gas runway accretion. It will also be critical to understand from
modeling whether the elemental composition at the bottom of my modeling
domain at 1000~bars is representative of the entire gas envelope
or whether the initial formation of a metal rich atmosphere and subsequent
runaway gas accretion can lead to incompletely mixed, stratified gas
envelopes.

Eventually, exoplanet observations may provide a statistical sample
of precise C/O ratios in giant planet envelopes ranging from hot Jupiter
to wide separation directly imaged giant planets. The Juno mission
--- expected to arrive in 2016 --- will obtain the first reliable
C/O measurement of for Jupiter. Meanwhile, disk observation may provide
a better understanding of the elemental abundances in the protoplanetary
gas and dust/ices. Together, observations of protoplanetary disks
and evolved planets provide the initial conditions as well as outcome
of planet formation, providing us with the unique opportunity to develop
a consistent picture of the formation of giant planets.

\section{Acknowledgements}

I would particularly like to thank Julie Moses for valuable discussions
on photochemistry and providing me access to her chemical reactions
list and detailed model outputs for model comparison and validation.
I would like to thank Caroline Morley, Mark Marley, and Jonathan Fortney
for providing me with detailed information of their atmospheric models
which enables me with the opportunity to compare and validate the
SCARLET forward model. I thank Heather Knutson for valuable feedback
on the manuscript. 

\FloatBarrier

\appendix{}

\section{Efficient Computation of Line-by-Line Radiative-Convective Equilibrium}

Converging to radiative-convective equilibrium generally requires
thousands of radiative transfer calculations of the entire atmospheric
column; hence previous models have either approximated the T-p profiles
based on analytical solution for (semi-)gray opacities \citep{guillot_radiative_2010,parmentier_non-grey_2014}
or by assuming correlated opacities across the entire atmospheric
column \citep{burrows_nongray_1997,marley_clouds_2002,fortney_synthetic_2008,morley_quantitatively_2013}.
Here, I present a novel numerical technique to speed up repeated line-by-line
radiative transfer calculations by orders of magnitudes --- enabling
the efficient computation of self-consistent T-p profiles for any
atmospheric composition based on line-by-line equivalent radiative
transfer. The key element of the new numerical technique is a three-dimensional
radiative-connectivity array, $L$, describing the radiative link
between any two layers in the atmosphere as a function of wavelengths.
Once $L$ is computed the T-p profile is computed using only a few
hundred of spectral points, with virtually no loss in the precision
compared to the full high-resolution line-by-line radiative calculations
with millions of spectral points. 

The basic concept of any radiative-convective model is to iterate
the T-p profile until the radiative downward flux $F^{\downarrow}$
due to direct star light and infrared reemission is matched by the
upward emission flux in the atmosphere. In the process, atmospheric
layers with temperature gradients exceeding the adiabatic lapse rate
$-\frac{dT}{dz}>\Gamma=\frac{g}{C_{p}}$ are declared convective and
adjusted to the adiabatic lapse rate. For purely absorbing atmospheres,
the exact downward and upward fluxes are 

\begin{equation}
F_{i}^{\downarrow}\left(\tau_{i}\right)=\intop_{0}^{\infty}F_{*,\lambda}e^{-\tau_{i}/\mu}d\lambda+\intop_{0}^{\infty}\intop_{0}^{\tau}\pi B_{\lambda}\left(\tau\right)\frac{d}{d\tau}T_{\lambda}^{f}\left(\tau_{i}-\tau\right)d\tau d\lambda\label{eq:Fdown}
\end{equation}

\begin{equation}
F_{i}^{\uparrow}\left(\tau_{i}\right)=\intop_{0}^{\infty}\pi B_{\lambda}\left(\tau_{*}\right)T_{\lambda}^{f}\left(\tau_{*}-\tau_{i}\right)d\lambda+\intop_{0}^{\infty}\intop_{\tau_{*}}^{\tau}\pi B_{\lambda}\left(\tau\right)\frac{d}{d\tau}T_{\lambda}^{f}\left(\tau_{i}-\tau\right)d\tau d\lambda\label{eq:Fup}
\end{equation}

where the optical depth $\tau_{i}$ is the vertical coordinate, $\tau^{*}$
is the optical depth of the planet's surface, $B_{\lambda}\left(\tau\right)$
is the Planck function at wavelength $\lambda$ for the atmospheric
temperature at optical depth $\tau$, and $F_{*,\lambda}$ is the
stellar irradiation at the top of the atmosphere. The term $T_{\lambda}^{f}\left(\tau_{i}-\tau\right)$
is the slab transmittance 

\begin{equation}
T_{\lambda}^{f}\left(\tau_{i}-\tau\right)=2\intop_{0}^{1}e^{-(\tau_{i}-\tau)/\mu}\mu d\mu\label{eq:slabtransmittance}
\end{equation}

between $\tau_{i}$ and $\tau$, where $\mu=\cos\theta$ is the cosine
of the inclination towards the upward normal. Conceptually, the two
terms in Equations \ref{eq:Fdown} are the contribution from direct
stellar irradiation (1st term) and the downward infrared emission
from atmospheric layers above $\tau$ (2nd term). The upwards flux
(Equation \ref{eq:Fup}) is composed of the infrared emission from
the planetary surface or deep interior (1st term) and the infrared
emission from the atmospheric layers below $\tau_{i}$. Iterating
towards convergence, i.e. $F^{\downarrow}\left(\tau_{i}\right)=F^{\downarrow}\left(\tau_{i}\right)$
for all $\tau_{i}$, is computationally extremely costly because each
iteration towards convergence requires the evaluation of the double
integrals in Equations \ref{eq:Fdown} and \ref{eq:Fup}, where capturing
the high-frequency variations in the molecular line opacities requires
millions of spectral points.

The key concept of the new line-by-line radiative-convective model
is to break the wavelength integration into parts for which the smoothly
varying Planck function, $B_{\lambda}$$\left(\tau\right)$, is approximate
constant, pulling $B_{\lambda}\left(\tau\right)$ out of the wavelength
integrals, and storing the piecewise integrated transmission terms
in a three-dimensional radiative-connectivity array, $L$, that can
be reused for many iterations till the converged temperature pressure
profile is obtained. 

As an example, the second term in Equation \ref{eq:Fup} in discretized
atmospheres is

\begin{equation}
F_{i,\mathrm{IR}}^{\uparrow}\left(\tau_{i}\right)=\intop_{0}^{\infty}\sum_{j=i}^{N}\pi B\left(T_{j},\lambda\right)\left[T_{\lambda}^{f}\left(\tau_{i}-\tau_{j}\right)-T_{\lambda}^{f}\left(\tau_{i}-\tau_{j+1}\right)\right]d\lambda,\label{eq:discretized}
\end{equation}

where the upward IR flux at level $i$ has contributions from all
layers $j=i\ldots N$ below the layer $i$, and $B_{\lambda}\left(T_{j}\right)$
is the Planck function at the temperature $T_{j}$ in layer $j$.
Repeated computations of Equation \ref{eq:discretized} can accelerated
by orders of magnitude by rewriting Equation \ref{eq:discretized}
as 

\begin{equation}
F_{i,\mathrm{IR}}^{\uparrow}\left(\tau_{i}\right)=\pi\sum_{j=i}^{n}\sum_{k=1}^{n_{k}}B\left(T_{j},\lambda_{k+1/2}\right)C_{ijk},
\end{equation}

where
\begin{equation}
C_{ijk}=\intop_{\lambda_{k}}^{\lambda_{k+1}}\left[T_{\lambda}^{f}\left(\tau_{i}-\tau_{j}\right)-T_{\lambda}^{f}\left(\tau_{i}-\tau_{j+1}\right)\right]d\lambda.
\end{equation}

forms the three-dimensional radiative-connectivity array, $C$ and
, $n_{k}$ is the number of spectral elements saved in $C$. Hundreds
of spectral elements suffice to perform a line-by-line equivalent
evaluation because $B\left(\lambda,T\right)$ varies slowly as a function
of $\lambda$ and no loss information occurs by precomputing $C_{ijk}$.
The other term terms in Equations \ref{eq:Fdown} and \ref{eq:Fup}
can be rewritten in a similar way. 

Once all elements $C_{ijk}$ are computed, the evaluation of Equations
\ref{eq:Fdown} and \ref{eq:Fup} is trivial and convergence to radiative-convective
equilibrium is achieved within seconds on a modern computer. Since
molecular opacities are temperature dependent, $C_{ijk}$ needs to
be repeated 3-4 times, until the model atmosphere is in radiative-convective
equilibrium and the opacities correspond to the final temperature
profile. Overall, the convergence to radiative equilibrium of 2-3
orders of magnitude faster performing line-by-line computations at
each iterative step toward radiative-convective equilibrium.

\bibliographystyle{apj}
\addcontentsline{toc}{section}{\refname}\bibliography{MyLibrary_v20140113}

\end{document}